\documentclass[referee,a4paper,12pt,traditabstract]{swsc} 

\usepackage{graphicx}
\usepackage{txfonts}
\usepackage{subfigure}
\usepackage{epstopdf}
\usepackage[displaymath,mathlines]{lineno}
\usepackage[authoryear,round]{natbib}
\usepackage{url}

\usepackage[T1]{fontenc}
\usepackage{pgffor}

\begin{document}


   \title{Immediate and delayed responses of power lines and transformers in the Czech electric power grid to geomagnetic storms}

   \titlerunning{Responses of Czech electric power grid equipment to geomagnetic storms}

   \authorrunning{\v{S}vanda et al.}

  \author{Michal \v{S}vanda \inst{1,2}\fnmsep\thanks{Corresponding author: \email{michal@astronomie.cz}}
		   \and
		   Didier Mourenas \inst{3}
		   \and
		   Karla \v{Z}ertov\'a\inst{4}
		   \and
		   Tatiana V\'ybo\v{s}\v{t}okov\'a\inst{5}
		   }

  \institute{Astronomical Institute of the Czech Academy of Sciences, CZ-25165 Ond\v{r}ejov, Czech Republic
             \and
			 Astronomical Institute, Charles University, CZ-18000 Praha, Czech Republic
			 \and
			 CEA, DAM, DIF, F-91297, Arpajon, France
			 \and
			 Gymn\'azium Ji\v{r}\'iho Ortena, CZ-284 80 Kutn\'a Hora, Czech Republic
			 \and
			 Department of Surface and Plasma Science, Charles University, CZ-18000 Praha, Czech Republic
			 }


 
  \abstract
{Eruptive events of solar activity often trigger abrupt variations of the geomagnetic field. Through the induction of electric currents, human infrastructures are also affected, namely the equipment of electric power transmission networks. It was shown in past studies that the rate of power-grid anomalies may increase after an exposure to strong geomagnetically induced currents. We search for a rapid response of devices in the Czech electric distribution grid to disturbed days of high geomagnetic activity. Such disturbed days are described either by the cumulative storm-time $Dst$ or $d(\textit{SYM-H})/dt$ low-latitude indices mainly influenced by ring current variations, by the cumulative $AE$ high-latitude index measuring substorm-related auroral current variations, or by the cumulative $ap$ mid-latitude index measuring both ring and auroral current variations. We use superposed epoch analysis to identify possible increases of anomaly rates during and after such disturbed days. We show that in the case of abundant series of anomalies on power lines, the anomaly rate increases significantly immediately (within 1 day) after the onset of geomagnetic storms. In the case of transformers, the increase of the anomaly rate is generally delayed by 2--3 days. We also find that transformers and some electric substations seem to be sensitive to a prolonged exposure to substorms, with a delayed increase of anomalies. Overall, we show that in the 5-day period following the commencement of geomagnetic activity there is an approximately 5--10\% increase in the recorded anomalies in the Czech power grid and thus this fraction of anomalies is probably related to an exposure to GICs.}


   \maketitle


\section{Introduction}

The Sun is a magnetically active star, filling the interplanetary space with a stream of charged particles called the solar wind \citep[see e.g. a recent review by][]{Verscharen2019}. The solar wind properties are far from being homogeneous, with strong variations in temperature, density, or interplanetary magnetic field observed in connection with various phenomena of solar activity. The main drivers of strong disturbances of the solar wind are coronal mass ejections (CMEs), the fast--slow solar wind interaction on the borders of corotating interaction regions, and fast-wind outflows from coronal holes. Solar-wind disturbances may ultimately interact with Earth's magnetosphere, thereby triggering geomagnetic activity. 

As first proposed by \citet{Dungey1961}, the dynamic pressure exerted by the solar wind on the magnetosphere can trigger magnetic reconnection, opening dayside dipolar geomagnetic field lines. The solar wind then transports this magnetic field to the nightside, forming a long tail behind the Earth. This transfer of magnetic flux and the resulting reconfiguration of the magnetosphere eventually leads to nightside magnetic reconnection, returning flux to the dayside in various phenomenological response modes that depend on the disturbance level \citep{Dungey1961, Kepko2015}. However, a common characteristic of all such response modes is the formation of a current wedge system \citep{Kepko2015, McPherron&Chu17:ssr}. A fraction of the tail current along geomagnetic field lines is then temporarily diverted through the ionosphere, allowing a closure of the current wedge and causing perturbations in the auroral zone and at middle latitudes \citep{McPherron&Chu17:ssr}. 

Both substorms and geomagnetic storms give rise to a current wedge, plasma sheet inward convection by inductive electric fields, and energetic particle injections \citep{Ganushkina2017, Kepko2015, McPherron&Chu17:ssr, Thomsen2004}. However, the current wedge has generally a more limited temporal extent during substorms than during storms, which frequently last for days \citep{Ganushkina2017, Kepko2015}. Substorms are one of the key dynamical processes occurring during storms, but isolated substorms also occur outside storms \citep{Viljanen2006, Turnbull2009}. During storms (mainly caused by strong interactions between CMEs and the magnetosphere), a stronger buildup of the inner ring current (a westward current of ions roughly $\sim2-4$ Earth radii above the equator) is provided by a deeper inward transport of charged particles from the plasma sheet, leading to a significant and prolonged decrease of the geomagnetic field \citep{Ganushkina2017}.

All these ionospheric and magnetospheric currents, and the related field-aligned currents, can cause important geomagnetic field variations during periods of rapidly evolving solar wind dynamic pressure \citep{Gonzalez94, Lakhina2016, McPherron&Chu17:ssr, Kappenman2005, tsurutani2009brief}. This realization has led to the traditional concept of disturbed days: days of smooth and regular geomagnetic field variations have been called {\it quiet days}, whereas days of stronger and irregular variations have been called {\it disturbed days} \citep{ChapmanBartels1940}.

Geomagnetically induced currents (GICs) in the ground are due to strong variations $dH/dt$ of the horizontal component $H$ of the geomagnetic field over typical time scales of $\sim 10-1000$ seconds during disturbed days \citep{Carter2015, Kappenman2003, Kataoka2008, Pokhrel2018,Zhang2016}. Substorms generally produce the largest $dH/dt$ at high and mid-latitudes during periods of fast solar wind and have caused many of the major GIC impacts during large storms -- e.g., the Quebec voltage collapse on 13 March 1989 was triggered by a first substorm, while two later substorms tripped out transformers in the UK \citep{Boteler2019}. $dH/dt$ was found to be twice smaller in general during non-storm substorms than during storm-related substorms, possibly due to an additional input from ring current variations during storms \citep{Viljanen2006, Turnbull2009}. Other important sources of $dH/dt$ during geomagnetic storms include sudden commencements (the shock compression of the magnetosphere when a fast CME impacts the magnetosphere at the start of a storm, leading to an increase of Chapman-Ferraro currents at the dayside magnetopause; e.g., see \citealt{Kikuchi2001}) and rapid variations of the ring current, through its role in the generation of Region 2 field-aligned currents \citep{Ganushkina2017}. Sudden commencements have a large $dH/dt$ because of their shock-like nature, while rapid increases of ring current energy density following large scale injection or inward convection of energetic charged particles coming from the plasma sheet can also produce large $dH/dt$ \citep{Kappenman2003, Kappenman2005, Kataoka2008}.

GICs propagate through conducting regions in the ground and water, but also in the grounded conductors. The presence of GICs in the electric power grid can cause various kinds of damage. GICs are quasi-DC currents that can lead to half-cycle saturation and drive a transformer response into a non-linear regime. This poses a risk for transformers by producing high pulses of magnetizing current, a local heating (also vibration) within the transformer \citep{Gaunt2014}, and the generation of AC harmonics that propagate out into the power network, where they can disrupt the operations of various devices \citep{Kappenman2007, Molinski2002}. In particular, the propagation of harmonics in the power grid during half-cycle saturation can distort the electrical current waveform, eventually triggering a detrimental reaction of protective relays connecting power lines, or leading to a disruption of other devices attached to these lines. 
 
GICs identified by fast variations of the geomagnetic field have been linked with various power grid failures \citep{schrijver2013disturbances}, eventually leading to power grid disruptions \citep{Kappenman2007, pirjola2000geomagnetically, Pulkkinen2017, schrijver2013disturbances}. Although high latitude regions are more at risk from GICs, middle and low latitude regions may also be impacted by significant GICs \citep{bailey2017, Carter2015, gaunt2007, Lotz2017, Marshall2012, Torta2012, Tozzi2019, Wang2015, Watari2009, Zhang2016, Zois2013}. 

A first study of anomalies in the Czech power grid as a function of geomagnetic activity (defined by the $K$ index computed from the measurements of the Earth's magnetic field at a local magnetometer station near Budkov -- e.g., see \citealt{Mayaud1980,McPherron&Chu17:ssr}) has already identified some statistically significant increases of the rate of anomalies around month-long periods of higher geomagnetic activity than nearby periods of lower activity \citep{Vybostokova2018}. Nevertheless, the relationship between geomagnetic events and anomalies still remained somewhat loose. 

Accordingly, the main goal of the present paper is to better ascertain the existence of a tight relationship between power grid anomalies and geomagnetic storms, on the basis of the same data set. We shall discuss the physical mechanisms by which GICs may cause anomalies in power lines and transformers, and show that our statistical results are suggestive of a causal relationship based on those mechanisms. We shall also address the important and unanswered question of the time delay between moderate to large geomagnetic storms with minimum $Dst<-40$ nT \citep{Gonzalez94} and the actual occurrences of anomalies. For that purpose, we shall use Superposed Epoch Analysis to investigate the relative occurrence of GIC effects in the Czech power grid during disturbed days as compared with quiet days. Such disturbed days will be categorized using different time-integrated parameters of geomagnetic activity, related to the magnitude of temporal variations of the horizontal component of the geomagnetic field, which can induce detrimental currents in power lines. 

\section{Data sets}
In this study, we searched for a causal relation between two types of time series. The first series describing the daily anomaly rates in the Czech electric power-distribution grid, and the second serving as a proxy of disturbed days for the estimation of geomagnetically induced currents. 

\subsection{Logs of Anomalies}
The Czech Republic is a mid-latitude country (around $\sim 50^\circ$ geographic latitude and $\sim 45^\circ$ corrected geomagnetic latitude), where the effects of solar/geomagnetic activity on ground-based infrastructures is expected to be moderate at most. The modelled amplitudes of GICs during the Halloween storms in late October 2003 reached 1-minute peaks of about 60~A\footnote{Smi\v{c}kov\'a, A., Geomagnetically Induced Currents in the Czech Power Grid, BSc. thesis (supervisor \v{S}vanda, M.), Faculty of Electrical Engineering, Czech Technical University, 2019, available online \url{http://hdl.handle.net/10467/84988}.}. The country has a shape prolonged in the east--west direction (about 500 km length), whereas in the south--north direction it is about 280~km long from border to border. The spine of the electric power network is operated by the national operator \v{C}EPS, a.s., which maintains the very-high-voltage (400~kV and 220~kV) transmission network, and connects the Czech Republic with neighbouring countries. \v{C}EPS also maintains the key transformers and electrical substations in the transmission network. The area of the state is then split into three regions, where the electricity distribution is under the responsibility of the distribution operators. The southern part is maintained by E.ON Distribuce, a.s., the northern part by \v{C}EZ Distribuce, a.s., and the capital city of Prague is maintained by PREdistribuce, a.s. All three distributors maintain not only very-high-voltage (110 kV) and high-voltage (22 kV) power lines, but also connect the consumers via the low-voltage (400 V) electric power transmission network. 

All four above-mentioned power companies have agreed to provide us their maintenance logs. The datasets used in this study are exactly the same datasets already used in the study by \cite{Vybostokova2018}. Thus, we refer the reader to section 3.2 of this previous paper for a more detailed description of the datasets. By mutual non-disclosure agreement with the data providers, the datasets were anonymised (by removing the information about the power-company name, and also by changing the calendar date to a day number) and must be presented as such. The total time span is 12 years, but the span of individual maintenance logs provided by the operators is shorter, varying between 6 to 10 years. 

We only briefly recall that the obtained logs were cleaned from events that were obviously not related to variations of geomagnetic activity. From these logs, we keep only the dates when the events occurred and did not consider any other details. These inhomogeneous datasets (the log entries were provided by different individuals with varying levels of details and quality of the event description) were split into twelve subsets D1--D12, which were investigated separately. Each sub-dataset was selected so that it contained only events occurring on devices of a similar type and/or with the same voltage level and were recorded by the same operating company. The dataset descriptions are briefly summarised in Table~\ref{tab:datasets}. 

\begin{table}[ht]
    \caption{Datasets analysed in this study. This is a reduced version of Table~1 in \cite{Vybostokova2018}.}
    \label{tab:datasets}
    \centering
    \begin{tabular}{l|lll}
      {\bf Dataset} & {\bf Voltage level} & {\bf Type} & {\bf Span}\\
      {\bf ID} & {\bf } & {\bf } & {\bf } \\
      \hline
      D1 & very high voltage & equipment: transformers, & 9 years\\
         &                   & electrical substations& \\
      D2 & high voltage & equipment & 6 years \\
      D3 & very high voltage & equipment & 6 years\\
      D4 & high and low voltage & power lines & 7 years\\
      D5 & high and low voltage & equipment and power lines & 7 years\\
      D6 & high and low voltage & equipment & 7 years \\
      D7 & very high voltage & power lines & 10 years \\
      D8 & high voltage & transformers & 10 years \\
      D9 & very high voltage & transformers& 10 years \\
      D10 & very high and high voltage & electrical substations& 10 years \\
      D11 & very high voltage & power lines& 10 years  \\
      D12 & high voltage & power lines & 10 years  \\
    \end{tabular}
\end{table}

\subsection{Geomagnetic Indices and Parameters used for GIC Estimation}

Various parameters have been considered to estimate the effects of geomagnetic activity on power grids \citep{schrijver2013disturbances}. GICs are due to strong variations $dH/dt$ over typical time scales of $\sim 10-1000$ seconds \citep{Kappenman2003}. There are two sources of such large $dH/dt$ at low and middle latitudes: (i) sudden impulses (SI), also called sudden commencements (SC) when they are followed by a storm caused by the shock preceding a fast CME, and (ii) the growth and decay of the ring current during a magnetic storm. Substorm-related disturbances are mostly limited to high and middle latitudes, whereas disturbances caused by ring current changes generally affect mainly middle and low latitudes. 
Statistically, periods of stronger cumulative effects of GICs in a power grid are therefore expected to correspond to {\it disturbed days} of elevated geomagnetic activity \citep{ChapmanBartels1940}. In the present study, we shall use various cumulative (time-integrated) parameters based on different magnetic indices to categorize such disturbed days, and we shall investigate the relative occurrence of GIC effects during such disturbed days as compared with quiet days.

An appropriate quantity to estimate GICs at low latitudes is $d(\textit{SYM-H})/dt$, which directly provides a (longitudinally averaged) measure of the 1-minute $dH/dt$ due to ring current variations that drive GICs there \citep{Carter2015, Kappenman2003, Zhang2016}. Indeed, the $\textit{SYM-H}$ index is essentially similar to the hourly $Dst$ storm time index, but measured on 1-minute time scales -- that is, it provides the disturbance $\Delta H = H - H_{\rm quiet}$ of the horizontal component of the magnetic field as compared to its quiet-time level, longitudinally averaged based on ground magnetometer measurements at different low latitude magnetometer stations \citep{Mayaud1980}. 

Several studies have demonstrated the existence of significant correlations between GICs or electric grid failures and times of large $d(\textit{SYM-H})/dt$ at low to middle latitudes during geomagnetic storms, although $d(\textit{SYM-H})/dt$ is often inappropriate during strong substorms \citep{Carter2015, Wang2015, Zhang2016}. \cite{Carter2015} have further shown that the actual $dH/dt$ at middle latitudes due to SI/SCs can be a factor $\sim 2-3$ larger on the dayside than $d(\textit{SYM-H})/dt$, potentially allowing GIC effects even during geomagnetic events with relatively small $d(\textit{SYM-H})/dt$. We checked that $dH/dt$ at the Czech magnetometer station of Budkov can also be sometimes $>2-3$ times larger than $d(\textit{SYM-H})/dt$ during SI/SCs. \cite{Viljanen2014} have noticed the presence of a European region of low underground conductivity stretching from France through Czech Republic to Hungary that could favor significant GICs at middle latitudes. \cite{Gil2019} have shown the presence of GICs during a few selected storms in Poland, while \cite{Tozzi2019} have found that non-negligible GICs could exist even down to northern Italy. \cite{Wang2015} have further emphasized that cumulative GICs in a nuclear plant transformer during a long-duration geomagnetic event could sometimes be more harmful than short events, due to the longer cumulated time of transformer heating. 

Accordingly, we consider here the $Int(d(\textit{SYM-H})/dt)$ parameter to categorize disturbed days of expected significant GIC impacts on power grids. $Int(d(\textit{SYM-H})/dt)$ is calculated over each day, as the sum of all 1-minute $\vert d(\textit{SYM-H})/dt\vert$ values (in nT/min) obtained during times when $\textit{SYM-H}$ remains smaller than some threshold. The selected threshold (varying from $-50$ nT to $-25$ nT) should ensure that only geomagnetic storm periods are considered \citep{Gonzalez94}. This $Int(d(\textit{SYM-H})/dt)$ parameter allows, in principle, to take into account the immediate effects on power grids caused by large individual $\vert dH/dt\vert$ due to ring current variations, as well as the more delayed, cumulative effects potentially caused by prolonged periods of moderate to significant $\vert dH/dt\vert$ levels \citep{Carter2015, Wang2015, Zhang2016} -- although large individual $\vert dH/dt\vert$ during strong substorms will need other indices such as $AE$ or $ap$ to take them into account (see below).

Other works have suggested that the mean or cumulative $Dst$ during storm main phase should be good indicators of long duration GICs, because larger and steeper decreases of $Dst$ correspond to stronger disturbances that should generally lead to larger $dH/dt$ at the relevant shorter time scales of $\sim 10-1000$ seconds \citep{Balan2014, Balan2016, Lotz2017}. Using observations in South Africa (at middle corrected geomagnetic latitudes $\sim 36^\circ-42^\circ$ not much lower than in the Czech Republic), \cite{Lotz2017} have demonstrated the existence of a linear relationship between the sum of induced electric fields recorded in the ground during geomagnetic storms and the integral of $\textit{SYM-H}$ (or $Dst$) values, suggesting that the cumulative $\textit{SYM-H}$ or $Dst$ could be used as good proxies for cumulated induced electric fields at middle corrected geomagnetic latitudes (although ring current effects are likely more important for GICs in South Africa than in the Czech Republic, where a more balanced mixture of ring current and substorm effects is present). They also noted that some effects might be present as long as $\textit{SYM-H}$ remained below $-20$ nT.

Therefore, we also consider the $IntDst$ parameter to categorize disturbed days of expected significant GICs in the Czech Republic \cite[e.g., see][]{Mourenas2018}. $IntDst$ (in nT$\cdot$hr) is calculated as a sum of hourly $\vert Dst\vert$ values. This summation starts when $Dst$ first becomes smaller than a threshold (taken between $-50$ nT and $-25$ nT as before) chosen to ensure that only storm periods are considered, and this summation ends when $Dst$ reaches its minimum value over the next 24 hours. Each $IntDst$ value is then assigned to the starting day of a given summation, with all integration periods strictly separated by construction. As a result, $IntDst$ is generally measured during storm main phase, where the effects on GICs are likely stronger \citep{Balan2014, Balan2016}, to provide a complementary metric to the $Int(d(\textit{SYM-H})/dt)$ metric calculated over each whole day without any consideration of storm phase.

While ring current variations during storms can be quantified by $Dst$ and $\textit{SYM-H}$ indices, the magnetic indices that provide a measure of magnetospheric and ionospheric current variations observed during strong substorms are $AE$, $AL$, $Kp$, or $ap$ \citep{Kamide1996, Mayaud1980, Mourenas2020, Thomsen2004}. The $ap$ index (as its logarithmic equivalent $Kp$) provides a global measure of the range of magnetic field variations at middle latitudes over 3-hour time scales, obtained by averaging measurements from different mid-latitude magnetometer stations spread in longitude \citep{Mayaud1980, Thomsen2004}. In contrast, the range indices $AE$ and $AL$ are measured at higher magnetic latitudes $>60^\circ$ inside the auroral region \citep{Mayaud1980, Kamide&Rostoker04}, and $AE$ saturates at high geomagnetic activity $am>150$ (with $am$ a mid-latitude index similar to $ap$) because the auroral oval then expands equatorward of the magnetometer stations measuring it \citep{Lockwood2019, Thomsen2004}. Therefore, $ap$ is probably more appropriate than $AE$ for quantifying the strength of time-integrated geomagnetic disturbances at middle (sub-auroral) geomagnetic latitudes than $AE$ \citep{Thomsen2004, Mourenas2020}.

Although $ap$ cannot provide an accurate ranking or quantification of the maximum $dH/dt$ values reached during the most disturbed events due to its intrinsic saturation at $Kp=9$ and its coarse 3-hour time resolution, it may still provide rough estimates during less extreme events with $Kp\sim3-7$ \citep{Kappenman2005}. Therefore, it is worth examining whether some time-integrated measure of $ap$ could still be used to simply categorize disturbed/quiet days of expected stronger occurrence/absence of GIC effects at middle latitudes, during a large series of medium (most frequent) to strong (more rare) time-integrated $ap$ events spread over 6 to 10 years. 

Accordingly, we shall consider in section~\ref{sect:intAP} a third parameter of geomagnetic activity, $IntAp$, corresponding to the daily maximum level of the integral of 3-hourly $ap$ values over a continuously active period of $ap\geq 15$ nT \citep{Mourenas2019, Mourenas2020}. This should allow to categorize disturbed days that include contributions to GICs from both (storm-time) ring current variations and strong substorms, usefully complementing the $Int(d(\textit{SYM-H})/dt)$ and $IntDst$ parameters. Indeed, $IntAp$ provides a rough estimate of the effects at middle latitudes of significant time-integrated $dH/dt$ disturbances due to substorms, which often do not reach the low latitudes where $\textit{SYM-H}$ and $Dst$ are measured.

In addition, we shall consider a fourth parameter, called $IntAE$, which is based on the high-latitude $AE$ auroral electrojet index \cite{Mayaud1980}. $IntAE$ is the daily maximum level of the integral of $AE$ calculated over the same period of continuously high $ap\geq15$ nT as $IntAp$ (generally corresponding to $AE>200$ nT), to ensure that the corresponding substorm-related magnetic disturbances effectively reach middle latitudes \citep{Mourenas2019, Mourenas2020}. $IntAE$ provides a measure of cumulative substorm-related disturbances, corresponding to continuous periods of auroral current variations roughly similar to High-Intensity Long-Duration Continuous $AE$ Activity (HILDCAA) events \citep{Tsurutani06}. 

\begin{figure}
    \centering
     \resizebox{0.9\textwidth}{!}{\rotatebox{-90}{\includegraphics{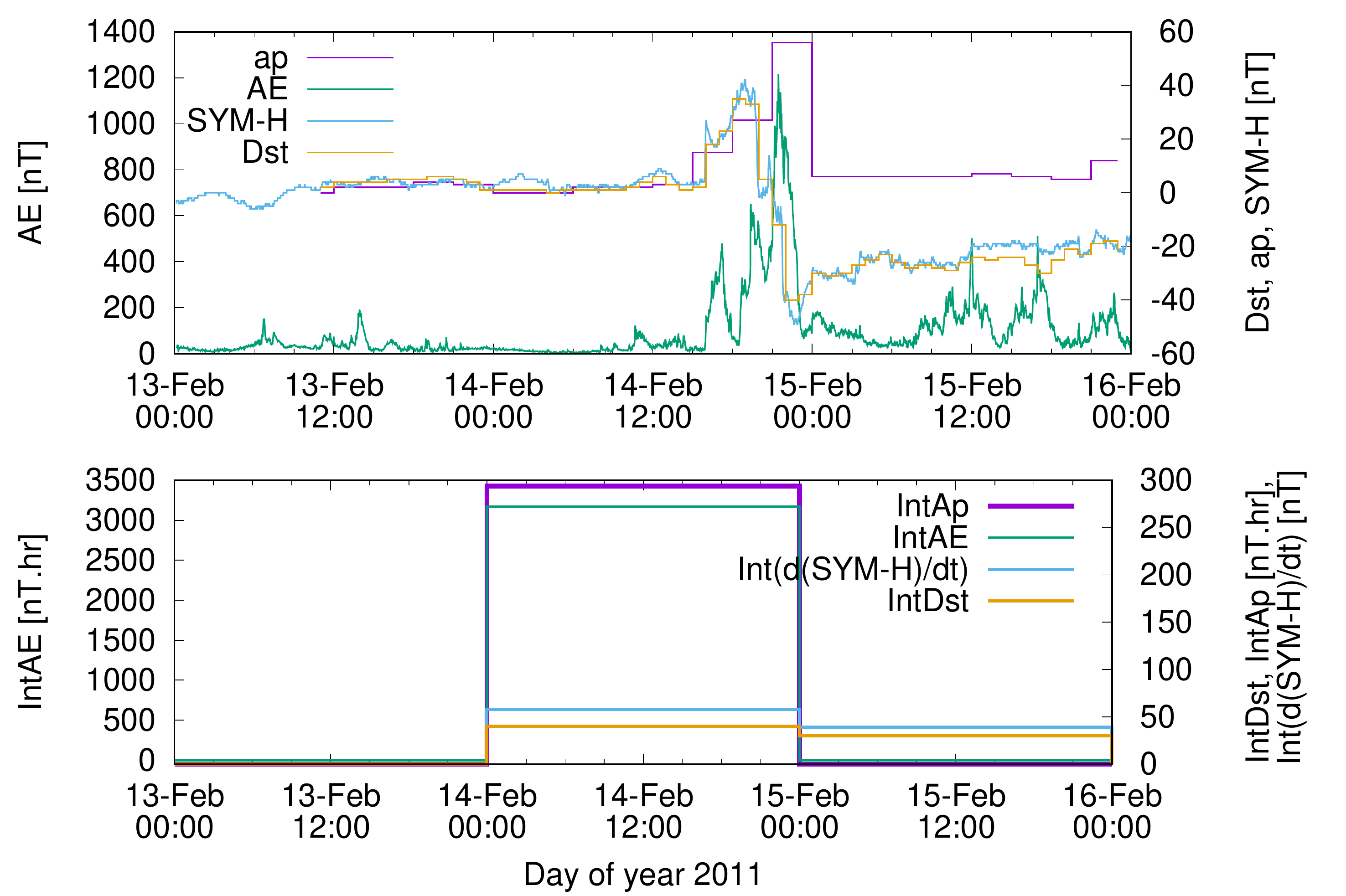}}}
     \caption{Upper panel: $\textit{SYM-H}$, $Dst$, $Ap$, and $AE$ indices during the 13-15 February 2011 geomagnetic event. Bottom panel: corresponding $Int(d(\textit{SYM-H})/dt)$ (in nT), and $IntDst$, $IntAp$, and $IntAE$ (in nT$\cdot$hr) cumulative parameters, calculated using thresholds $\textit{SYM-H}\leq-30$ nT, $Dst\leq-30$~nT, or $ap\geq15$ nT. }
    \label{fig:2011-02-13}
\end{figure}

These four cumulative metrics of disturbed days are displayed in Figure \ref{fig:2011-02-13} together with 1-min $\textit{SYM-H}$ and $AE$, hourly $Dst$, and 3-hourly $ap$, during a moderate geomagnetic storm on 14-15 February 2011 that reached a minimum $\textit{SYM-H}=-49$ nT and a minimum $Dst=-40$ nT on 14 February, with strong substorms (identified by peaks in $AE$ and $ap$) during storm sudden commencement and main phase, and with a very weak secondary minimum of $Dst$ reaching $-30$ nT on 15 February at 17 UT during a burst of $AE$ activity.

\section{Methods}

In the present follow-up study to the work by \cite{Vybostokova2018}, we search for a tighter relationship between power grid anomalies and geomagnetic storms, based on the same datasets of anomalies in the Czech power grid. We also address the important and as yet unanswered question of the time delay between geomagnetic events and the occurrences of anomalies.

Our working hypothesis is that {\it disturbed days} of high geomagnetic activity should cause an increase in daily rates of anomalies in the power distribution network as compared with {\it quiet days}. Accordingly, the daily anomaly rates should sharply peak within a few days (with some delay) after such disturbed days, and then decrease back to normal levels. This corresponds to a rapid response to GICs induced by substorms and storms, as observed for a few selected events -- e.g., see \cite{Gil2019, Wang2015}.

Unfortunately, in a mid-latitude country such as the Czech Republic, the effects of geomagnetic activity are expected to be weak. Consequently, an investigation of individual, moderate geomagnetic events is not expected to reveal a significant increase of anomalies, because such anomalies induced by geomagnetic activity (via GICs) will generally remain hidden among many other anomalies caused by various other effects. It is therefore imperative in our statistical analysis to find a way to reduce the importance of anomalies caused by other effects.  Note that our data series cover 6 to 10 years, each subset providing records of anomaly rates occurring during many separated disturbed days of high geomagnetic activity. Therefore, a feasible approach is to average over all these different events. The corresponding methodology is the \emph{Superposed Epoch Analysis}, widely used in astrophysics.

A Superposed Epoch Analysis \citep[SEA;][]{Chree1913} is a statistical technique used to reveal either periodicities within a time sequence, or to find a correlation between two time series. In the later case, the method proceeds in several steps.
\begin{enumerate}
    \item In the reference time series, occurrences of the repeated events are defined as key times (or epochs).
    \item Subsets are extracted from the second time series within some range around each key time.
    \item Subsets from each time series are superposed, synchronized at the same key time (Day 0), and averaged, allowing inter-comparisons.
\end{enumerate}
This methodology is known to efficiently enhance the ``signal'' (related variations in both series) with respect to ``noise'' (unrelated variations in both series), because the noise adds up incoherently, whereas the signal is reinforced by the superposition. 

Thus, we performed the SEA of geomagnetic activity defined by $Int(d(\textit{SYM-H})/dt)$ or $IntDst$ parameters. A range of event thresholds $\textit{SYM-H}$ (or $Dst$) $<-25$ nT to $-50$ nT was considered, to keep only periods corresponding to weak to large geomagnetic storms \citep{Gonzalez94} and to allow for the determination of the best thresholds on event strength. Other days were assigned a zero level of $Int(d(\textit{SYM-H})/dt)$ or $IntDst$. An important further requirement was that the 5-day period immediately preceding the start of a geomagnetic storm (Day 0 in the SEA) contained a zero level of the considered geomagnetic activity parameter (that is, all such quiet days must have $IntDst=0$ or $Int(d(\textit{SYM-H})/dt)=0$). This rather strict constraint should allow to better quantify the effect of geomagnetic storms on the power grid during {\it disturbed days} as compared with {\it quiet days}, at the expense of a slight reduction of the number of considered events. In a second step, we analyzed in more details these SEAs to determine as accurately as possible the time delay (after the start of a storm) that corresponds to the statistically most significant increase of anomalies, for each type of power grid equipment. 

\section{Results of Superposed Epoch Analysis}

A Superposed Epoch Analysis was performed based on $IntDst$ and $Int(d(\textit{SYM-H})/dt)$ parameters, considering successively thresholds $Dst<-25$ nT, $-30$ nT, $-40$ nT, and $-50$ nT, or $\textit{SYM-H}<-25$ nT, $-30$ nT, $-40$ nT, and $-50$ nT, to explore the dependence of power grid anomalies on the minimum strength of geomagnetic storms. The number of epochs considered in the SEAs of each reference series are given in Table~\ref{tab:epochs}.

\begin{table}[]
    \caption{The number of epochs considered in SEAs for various reference series. }
    \centering
    \begin{tabular}{ccl}
         {\bfseries Reference series} & {\bfseries Threshold} & {\bfseries \# of epochs}\\
         \hline
         $IntDst$ & $-50$~nT & 138 \\
         $IntDst$ & $-40$~nT & 172 \\
         $IntDst$ & $-30$~nT & 221 \\
         $IntDst$ & $-25$~nT & 222 \\
         $Int(d(\textit{SYM-H})/dt)$ & $-50$~nT & 154 \\
         $Int(d(\textit{SYM-H})/dt)$ & $-40$~nT & 191 \\
         $Int(d(\textit{SYM-H})/dt)$ & $-30$~nT & 218 \\
         $Int(d(\textit{SYM-H})/dt)$ & $-25$~nT & 231
    \end{tabular}
    \label{tab:epochs}
\end{table}

\begin{figure}
    \centering
    \resizebox{0.49\textwidth}{!}{\rotatebox{-90}{\includegraphics{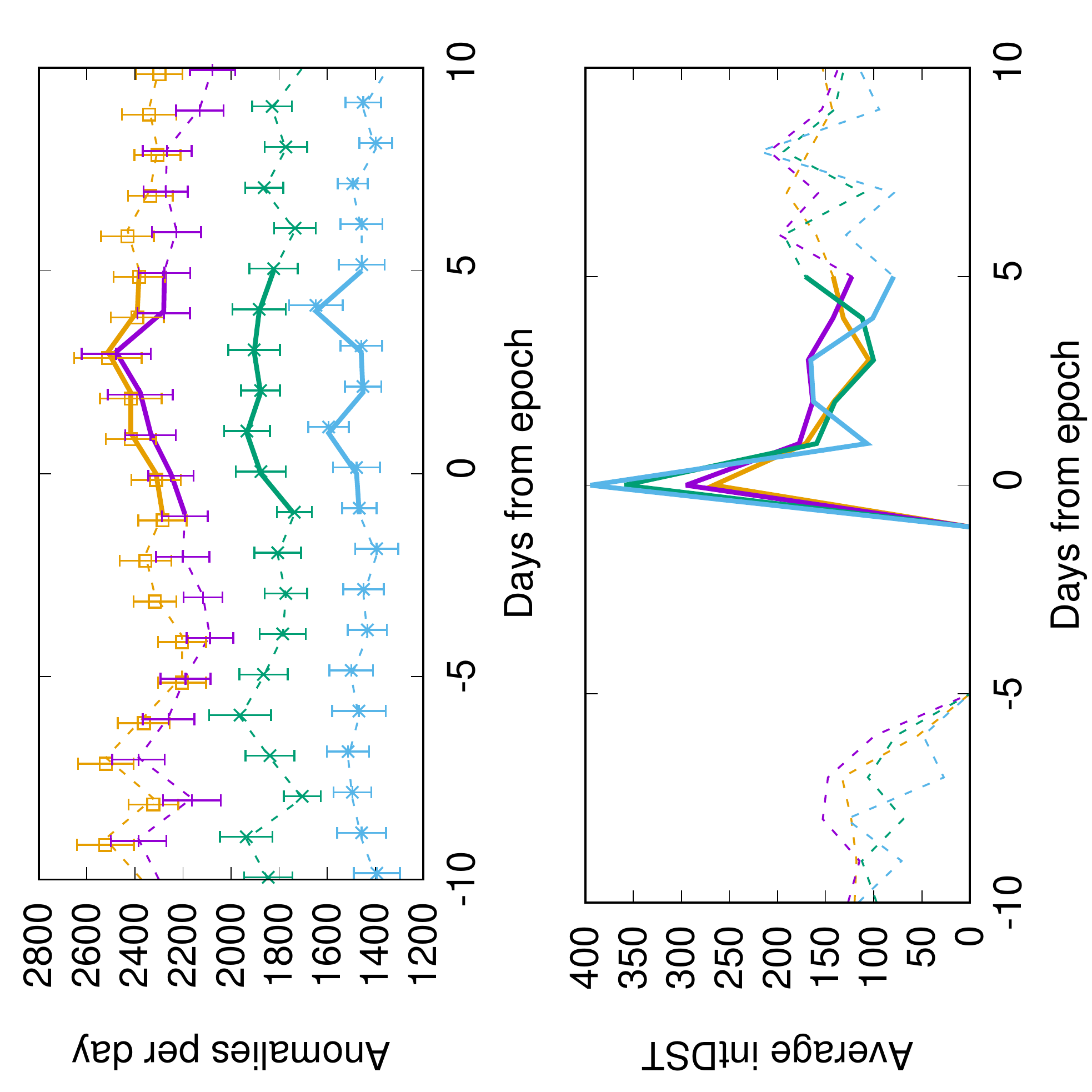}}}
    \resizebox{0.49\textwidth}{!}{\rotatebox{-90}{\includegraphics{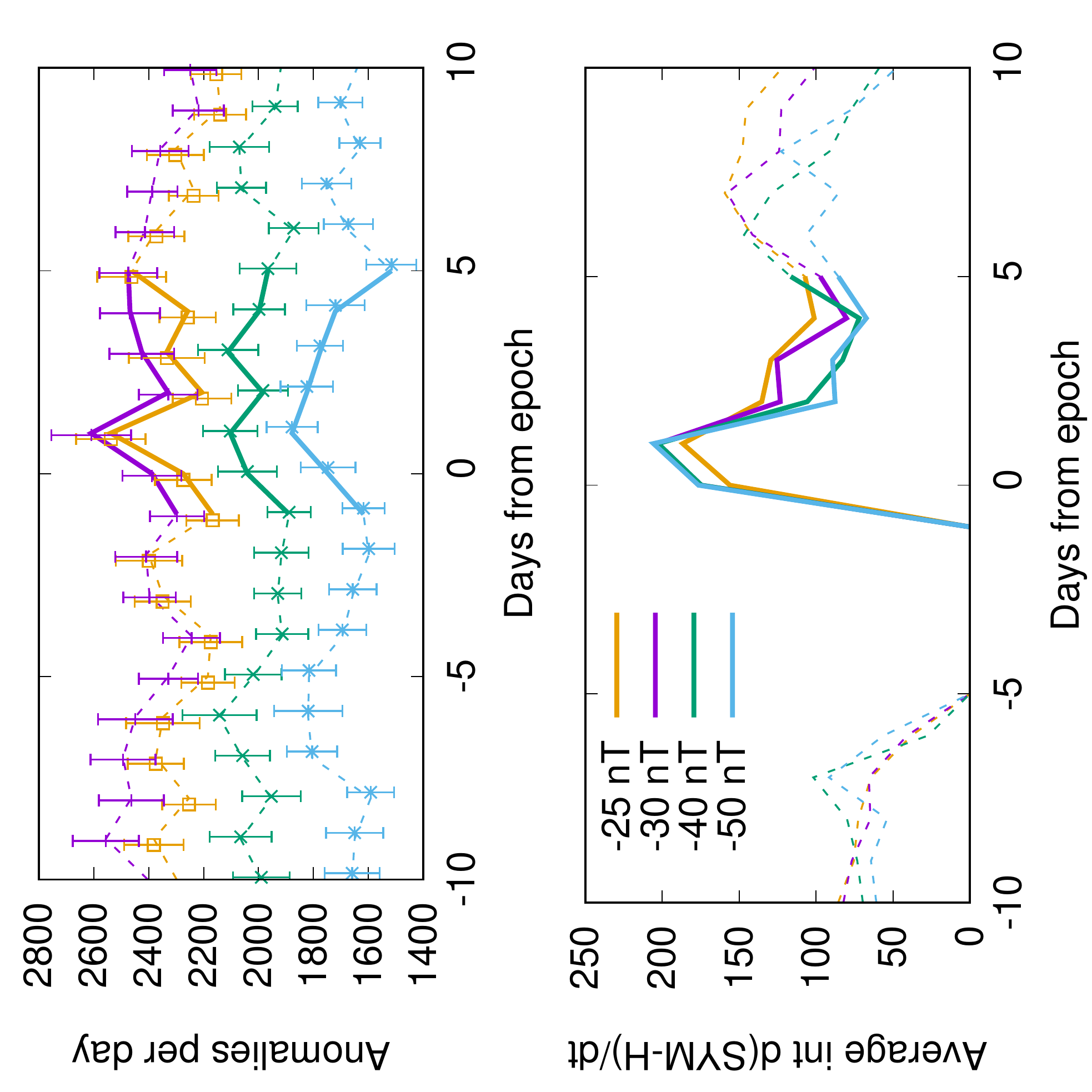}}}
    \caption{Plots of epoch-superposed daily numbers of anomalies in the D12 series, considering $IntDST$ (in nT$\cdot$hr, left) and $Int(d(\textit{SYM-H})/dt)$ (in nT, right) for different upper thresholds on $Dst$ and $\textit{SYM-H}$. Solid lines indicate the superposed anomaly rates (upper row) or geomagnetic activity in the reference time series (lower row) during Days~$-1$ to $+5$ from the epoch (Day~0), whereas dashed lines show the same quantities for the remaining days. Error bars show the one-standard-deviation half-widths. }
    \label{fig:SEA4}
\end{figure}

The SEAs obtained for $IntDst$ and $Int(d(\textit{SYM-H})/dt)$ both show a clear peak of geomagnetic activity at Day 0 and a sharp decrease on Day~1 for $IntDst$ or on Day~2 for $Int(d(\textit{SYM-H})/dt)$. The later decrease for $Int(d(\textit{SYM-H})/dt)$ is due to the presence of significant $d(\textit{SYM-H})/dt$ variations during the recovery phase of many storms stretching over at least 2 consecutive days, whereas $IntDst$ is generally calculated only during storm main phase. Fig. \ref{fig:SEA4} shows the SEAs obtained for the D12 series (power lines). Similar trends are found for other datasets concerning power lines. All the figures corresponding to the different series D1 to D12 are available in the online supplement as Figs.~\ref{fig:D1seas}-\ref{fig:D12seas}.

\subsection{Storm Effects: 5-day Periods After/Before Day 0}
\label{sect:5daysafterbefore}

Next, we compared the period of 5 {\it disturbed days} immediately following Day 0 (the day of peak storm activity) with the 5-day period immediately preceding Day 0 -- a preceding period of {\it quiet days} especially selected to have zero $IntDst$ or $Int(d(\textit{SYM-H})/dt)$ levels. This allows to directly check the impact of {\it disturbed days} of geomagnetic storms on power grid anomalies, as compared with {\it quiet days}. For the two time intervals, we summed the total number of registered anomalies in the superposed series for each data subset and computed the statistical significance of the differences using the standard binomial statistical test. We tested the null hypothesis that the number of anomalies recorded over quiet days is not different from the number of anomalies recorded over disturbed days, that is, the null hypothesis that the probability of recording anomalies is the same during quiet and disturbed days. Should the resulting $p$-value be smaller than the selected statistical threshold (usually 0.05 for single-bin tests), we reject the null hypothesis, thereby saying that the recorded differences are indeed statistically significant. 

\begin{table}[]
    \caption{Comparison of the number of power grid anomalies in the 5-day period prior to Day~0 $N_{-}$ and in the 5-day period after Day 0~$N_{+}$, together with $p$-values of the statistical significance of the differences. These values are given for different reference series involved in SEAs with varying thresholds.  }
    \centering
    $IntDst$
    \begin{tabular}{l|lll|lll|lll|lll}
    {\bfseries ID} & \multicolumn{3}{c|}{$<-25$~nT} & \multicolumn{3}{c|}{$<-30$~nT} & \multicolumn{3}{c|}{$<-40$~nT} & \multicolumn{3}{c}{$<-50$~nT}\\
    & $N_{-}$ & $N_{+}$ & $p$ & $N_{-}$ & $N_{+}$ & $p$ & $N_{-}$ & $N_{+}$ & $p$ & $N_{-}$ & $N_{+}$ & $p$\\
    \hline
    D1 &  60 & 59 & 1.0 &  54 & 52 &  0.92 &  35 & 33 &  0.90 &  29.0 &  36.0 & 0.46\\
    D2 & 100 & 115 & 0.34 &  109 & 137 & 0.08 &   94 & 112 & 0.24 &   82 & 94 &  0.41\\
    D3 & 17 & 17 & 1.0 &  20 & 23 & 0.76 &   16 &   22 &  0.42 &  18 & 12 & 0.36\\
    D4 &  58 & 38 & 0.05 & 52 & 43 & 0.41 &  45 & 46 & 1.0 & 38 & 40 & 0.91 \\
    D5 &  86 & 75 & 0.43 &  91 & 84 & 0.65 &  83 & 82 & 1.0 &  71 & 68 & 0.87\\
    D6 & 30 & 36 & 0.54 &  40 & 39 & 1.0 & 38& 37 & 1.0 &  34& 31& 0.80\\
    D7 &  134 & 132 & 0.95 &  143 & 137 & 0.77 &  115 & 120 & 0.79 &   98 & 105 & 0.67\\
    D8 & 968 & 955 & 0.78 &  892 & 922 & 0.50 &  710 & 760 & 0.20 &  562 & 586 & 0.50\\
    D9 & 105 & 102 & 0.89 &  95 & 112 & 0.27 &  70 & 67 & 0.86 &  44 & 53 & 0.42\\
    D10 & 14292 & 14338 & 0.79 &  13245 & 13477 & 0.16 &   10791 & 11047 & 0.08 &  8601 & 8764 & 0.22\\
    D11 & 415 & 494 & 0.01 &  403 & 476 & 0.02 &   302 & 387 & $<0.01$ & 247 & 297 & 0.04\\
    D12 & 11366 & 12118 & $<0.01$ & 10787 & 11748 & $<0.01$ &  8965 & 9421 & $<0.01$ &  7242 & 7606& $<0.01$    
    \end{tabular}

    \vskip5mm
    $Int(d(\textit{SYM-H})/dt)$
    \begin{tabular}{l|lll|lll|lll|lll}
    {\bfseries ID} & \multicolumn{3}{c|}{$<-25$~nT} & \multicolumn{3}{c|}{$<-30$~nT} & \multicolumn{3}{c|}{$<-40$~nT} & \multicolumn{3}{c}{$<-50$~nT}\\
    & $N_{-}$ & $N_{+}$ & $p$ & $N_{-}$ & $N_{+}$ & $p$ & $N_{-}$ & $N_{+}$ & $p$ & $N_{-}$ & $N_{+}$ & $p$\\
    \hline
    D1 & 59 & 56 & 0.85 &  59 & 58 & 1.0 &  43 & 47 & 0.75 &  32 & 37 & 0.63\\
    D2 & 98 & 98 & 1.0 &  104 & 110 & 0.73 &  101 & 121 & 0.20 &  93 & 107 & 0.36\\
    D3 & 20 & 15 & 0.50 &  20 & 16 & 0.62 &  15 & 20 & 0.50 &  17 & 18 & 1.0 \\
    D4 & 53 & 36 & 0.09 &  51 & 37 & 0.17 &  43 & 45 & 0.92 &  46 & 49 & 0.84\\
    D5 & 79 & 66 & 0.32 &  83 & 70 & 0.33 &  80 & 78 & 0.94 &   83 & 77 & 0.69\\
    D6 & 29 & 28 & 1.0 &  35 & 31 & 0.71 &  38 & 33 & 0.64 &  38 & 29 & 0.33\\
    D7 & 115 &118 & 0.90 &  137 & 127 & 0.58 &  116 & 122 & 0.75 &  119 & 118 & 1.0\\
    D8 & 964 & 936 & 0.54 &  1005 & 964 & 0.37 &  784 & 790 & 0.90 &  635 & 667 & 0.39\\
    D9 & 98 & 101 & 0.89 &  107 & 102 & 0.78 &  80 & 93 & 0.36 &  58 & 74 & 0.19\\
    D10 &  14220 & 14061 & 0.35 &  14594 & 14518 & 0.66 &  11951 & 11877 & 0.64 &  9702 & 9854 & 0.28\\
    D11 & 408 & 450 & 0.16 &  420 & 473 & 0.08 & 334 & 415 & $<0.01$ &  300 & 323 & 0.38\\
    D12 & 11273 & 11798 & $<0.01$ &  11675 & 12305 & $<0.01$ &  9669 & 10162 & $<0.01$ &  8385 & 8714 & 0.01 
    \end{tabular}
        
    \label{tab:pvalues}
\end{table}

The results, summarized in Table~\ref{tab:pvalues}, reveal a clear increase of anomalies during the period of 5 {\it disturbed days} following Day 0 as compared with the period of 5 {\it quiet days} preceding Day 0, for the two series D11 and D12 corresponding to power lines. The number of anomalies increases by 5\% for D12 and by 30\% for D11, with corresponding $p$-values always statistically significant ($<0.05$), for thresholds $<-30$ nT or $<-40$ nT -- except for $Int(d(\textit{SYM-H})/dt)$ and D11 for a threshold $<-30$ nT. Lower or higher thresholds usually lead to less statistically significant increases of anomalies, although not always -- e.g. for D11 and $IntDst$, the $<-25$ nT threshold gives a higher statistical significance. This means that moderate events with minimum $Dst$ or $\textit{SYM-H}$ near $-40$ nT have often a statistically detectable impact on anomaly rates, whereas weaker events do not. The same thresholds also lead to the highest peaks of anomalies after Day 0 in many other series. Finally, for D11 and D12, the $<-40$ nT thresholds lead to the smallest $p$-values ($<0.01$) for both $IntDst$ and $Int(d(\textit{SYM-H})/dt)$, as well as to the smallest $p$-values $<0.1-0.2$ for D8 and D10 when considering $IntDst$, and to the smallest or second smallest $p$-values $<0.2-0.36$ for D2 and D9 when considering $Int(d(\textit{SYM-H})/dt)$. Therefore, the thresholds $\textit{SYM-H}<-40$ nT and $Dst<-40$ nT are probably the most appropriate to detect statistically significant increases of anomalies related to geomagnetic storms. 

The weaker significance of results for higher thresholds $<-25$ nT agrees with previous observations from \cite{Lotz2017} that weaker events have little effects on induced electric fields. However, moderate $Dst$ or $\textit{SYM-H}$ geomagnetic disturbances in the range $-40$ nT to $-50$~nT are found to still have some impact on power lines. The weaker significance of results for lower thresholds $<-50$ nT is likely due to a combination of two different effects: (i) storms start slightly later when using a threshold $<-50$ nT than for higher thresholds $<-40$ nT or $<-30$ nT, meaning that the 5-day period preceding Day 0 can actually contain significant $dH/dt$ geomagnetic activity leading to some anomalies, and (ii) the $<-50$ nT threshold corresponds to a 30\% to 40\% smaller number of events than the $<-30$ nT threshold, decreasing the sensibility of the SEA to a potential slight increase of anomalies due to storms.

A detailed inspection of the SEAs of D12 lends further credence to the impact of geomagnetic storms on power lines. Indeed, for both $IntDst$ and $Int(d(\textit{SYM-H})/dt)$, the peaks of anomalies in the few days following Day 0 reach the highest daily levels of anomalies of the whole 21-day SEAs for $<-30$ nT to $<-50$ nT thresholds, the main increases of anomalies occurring from Day $+0$ to Day $+3$. For D11 and thresholds $<-30$ nT to $<-40$ nT, the 4-day period following Day 0 has also the highest number of anomalies of the whole 21-day SEA, while the 5-day interval preceding Day 0 has the lowest average number of anomalies of the whole SEA.

\subsection{Storm Effects: 3-day Periods Before/After Day 0 with Time Lags}

\label{sect:3days}
\begin{figure}
    \centering
    \includegraphics[width=\textwidth]{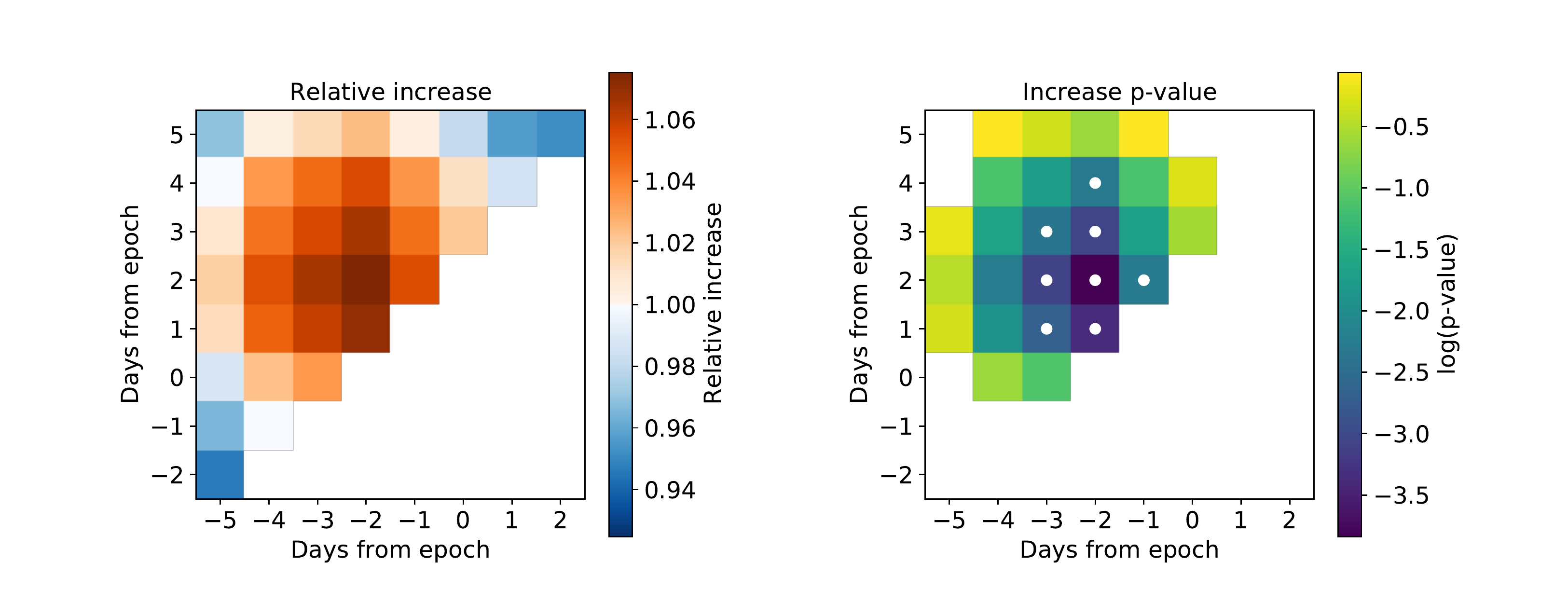}
    \includegraphics[width=\textwidth]{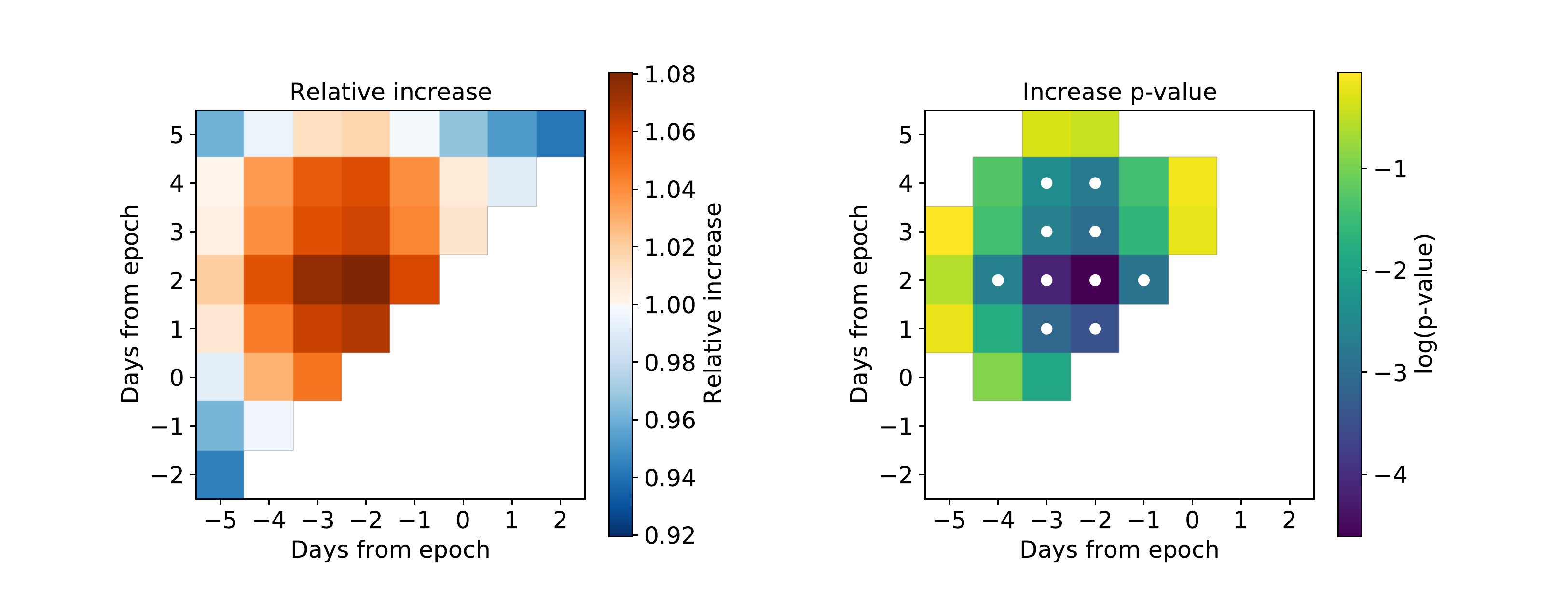}
    \caption{(left) Maps for D12 of increases (or decreases) of the number of anomalies as a function of the middle day of the first (abscissa) and second (ordinate) considered 3-day periods. (right) Maps of the corresponding $p$-values. The upper row is computed for the $IntDST$ reference series, whereas the lower row corresponds to the $Int(d(\textit{SYM-H})/dt)$ reference series. The $p$-values are evaluated only if there is an increase of anomaly rates in the second 3-day period as compared to the first 3-day period. Note the logarithmic scale of the plotted $p$-values: $p=0.0055$ (the adopted level of statistical significance for individual bins) corresponds to $\log p=-2.26$. Statistically significant bins are indicated by white dots. Blank bins are indicated by the white colour. }
    \label{fig:pvalues12}
\end{figure}

\begin{figure}
    \centering
    \includegraphics[width=\textwidth]{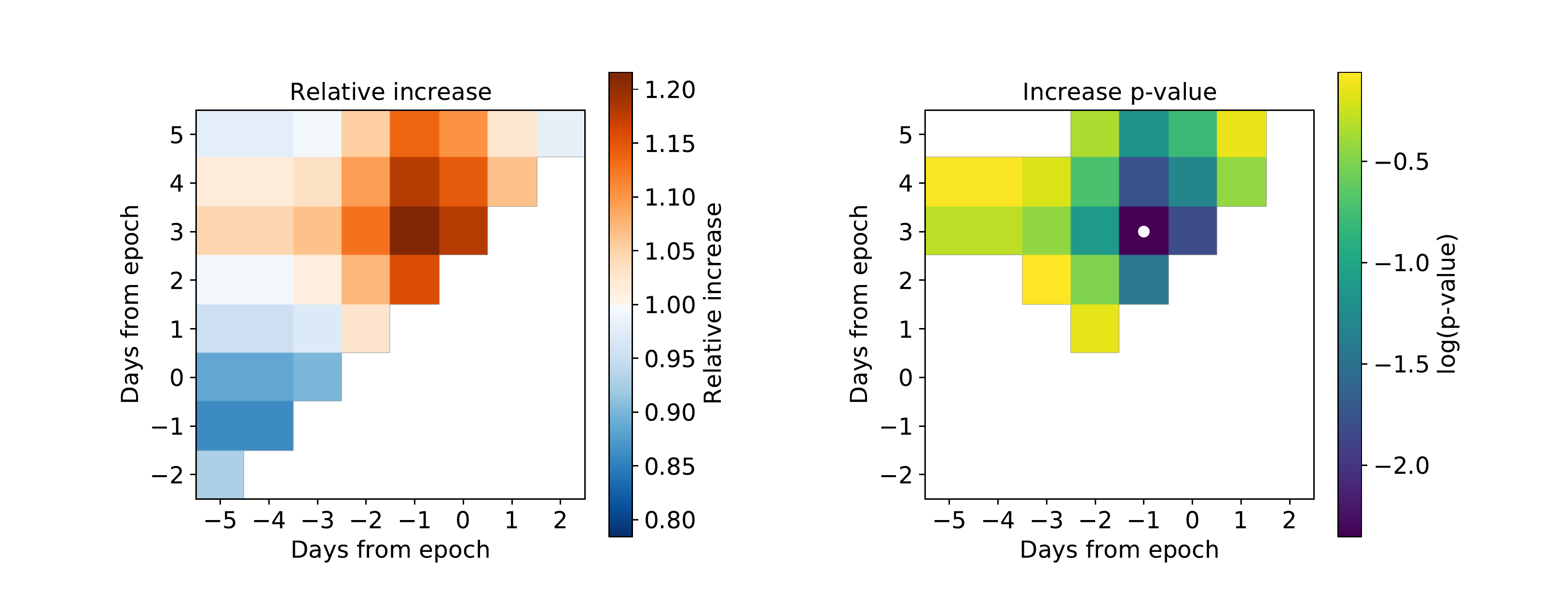}
    \includegraphics[width=\textwidth]{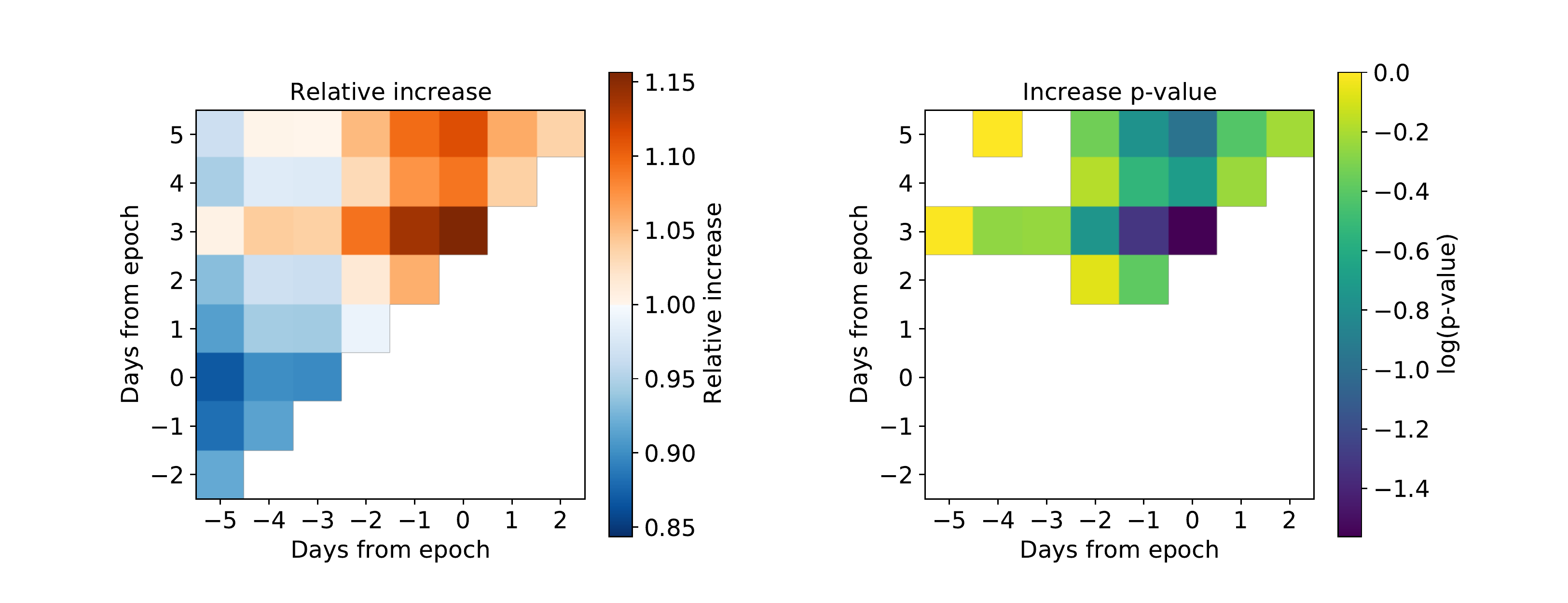}
    \caption{(left) Maps for D8 of increases (or decreases) of the number of anomalies as a function of the middle day of the first (abscissa) and second (ordinate) considered 3-day periods. (right) Maps of the corresponding $p$-values. The upper row is computed for the $IntDST$ reference series, whereas the lower row corresponds to the $Int(d(\textit{SYM-H})/dt)$ reference series. The $p$-values are evaluated only if there is an increase of anomaly rates in the second 3-day period as compared to the first 3-day period. Note the logarithmic scale of the plotted $p$-values: $p=0.0055$ (the adopted level of statistical significance for individual bins) corresponds to $\log p=-2.26$. Statistically significant bins are highlighted by white dots. Blank bins are indicated by the white colour.}
    \label{fig:pvalues8}
\end{figure}

Next, we examined in more details the SEAs performed based on $IntDst$ and $Int(d(\textit{SYM-H})/dt)$ parameters for thresholds $Dst<-40$ nT and $\textit{SYM-H}<-40$ nT. We considered two shorter 3-day periods, located before and after Day 0. We varied the time lag between them and calculated (as before for 5-day periods) the statistical significance of the difference in anomaly rates between these two periods. Considering shorter 3-day periods should help to determine more precisely the (statistically most significant) time delay between the start of a geomagnetic storm and the related increase of the number of anomalies. 

Fig. \ref{fig:pvalues12} for D12, Fig. \ref{fig:pvalues8} for D8, and Figs.~\ref{fig:D1ps}--\ref{fig:D12ps} in the online supplement for all other datasets, show two-dimensional maps of the increases (or decreases) of the number of anomalies as a function of the middle day of the first and second 3-day periods, together with maps of the corresponding $p$-values computed only for increases.

Let us examine these maps of $p$-values. For consistency with the procedure of estimation of the statistical significance adopted in Section~\ref{sect:5daysafterbefore}, we need to compare the number of anomalies over the same 5-day periods after and before Day 0. Accordingly, we must only consider the bins (representing 3-day periods) comprised between Days $-4$ and $-2$ (actually covering Days $-5$ to $-1$) for the period before Day 0, and the bins comprised between Days $+2$ and $+4$ (actually covering Days $+1$ to $+5$) for the period following Day 0. There are $3\times 3 = 9$ such bins. Finding only one bin with a $p$-value $\sim 0.05$ (corresponding to a 5\% probability to obtain an increase of anomalies by chance) among 9 bins is not anymore as statistically significant as before. Therefore, an individual bin (representing 3-day periods) is hereafter required to have a smaller $p$-value $\leq 0.05/9= 0.0055$ to be considered statistically significant.

In the case of the D12 dataset (power lines), there are six bins with $p$-values $< 0.0055$ for both $IntDst$ and $Int(d(\textit{SYM-H})/dt)$ in the considered square of $3\times 3$ bins centered on $(-3,+3)$ in Fig.~\ref{fig:pvalues12}, corresponding to a statistically significant increase of anomalies. A significant increase of anomalies is already observed over final 3-day periods centered on Day $+1$, as compared with initial 3-day periods centered on Days $-3$ and $-2$, indicating an immediate effect of geomagnetic storms on power lines. 

In the case of D8 (transformers), however, the three bins corresponding to increases of anomalies with the smallest $p$-values are found in Fig. \ref{fig:pvalues8} for final 3-day periods centered on Days $+3$ to $+4$, as compared with initial 3-day periods centered on Days $-1$ to $0$. Therefore, there is a clear time delay of $\sim 2-3$ days between a variation of $IntDst$ or $Int(d(\textit{SYM-H})/dt)$ and the corresponding variation of the number of anomalies in the D8 dataset. In such a situation, it is more appropriate to consider for D8 the square of $3\times 3 =9$ bins centered on $(-1,+3)$ in Fig. \ref{fig:pvalues12}. Inside this domain, one bin has a $p$-value $=0.0045 < 0.0055$ for $IntDst$ in Fig. \ref{fig:pvalues8}, indicating a statistically significant {\it delayed} increase of anomalies for D8. 

Overall, the results displayed in Figs. \ref{fig:pvalues12}-\ref{fig:pvalues8} and in Figs.~\ref{fig:D1ps}--\ref{fig:D12ps} therefore confirm the preceding results obtained for 5-day periods, but they further allow to determine the optimal time delays before a statistically significant increase of anomalies in different power grid equipment.

Most strikingly, a statistically highly significant increase of anomalies is found for D11--D12 (power lines) for both $IntDst$ and $Int(d(\textit{SYM-H})/dt)$ only $\sim 0-1$ day after Day 0, and as compared with all the preceding 3-day periods without storm activity (i.e., with $IntDst=0$ or $Int(d(\textit{SYM-H})/dt)=0$). Some less significant increases are also found for D4 (power lines as D11--D12) for $IntDst$. Such results imply an immediate effect of geomagnetic storms on power lines, already on Days 0 to $+1$. This looks quite realistic, because any effect of GICs on power lines (due to harmonics-related current waveform distortion leading to a detrimental reaction of protective relays or other devices connected to these lines) is likely to occur almost immediately.

Furthermore, Fig. \ref{fig:pvalues8} reveals the presence of a statistically significant {\it delayed} increase of anomalies for D8 (high voltage transformers) following geomagnetic storms when considering $IntDst$ (an increase is also present for $Int(d(\textit{SYM-H})/dt)$ but somewhat less significant), with a delay of $\sim 3$ days after Day 0. This strongly suggests the presence of some delayed effects of storm-time geomagnetic activity on transformers (note also that the lowest rates of anomalies are observed here on Days $-2$ to $0$, similarly corresponding to a delayed effect of the previous days of zero storm activity). Transformers may indeed be affected by GICs but still continue to operate for a while -- typically for a few days -- before actual problems ultimately show up and are registered in logs \citep[e.g.,][]{Wang2015}.

\subsection{Ring and Auroral Currents Effects: $IntAp$ parameter}
\label{sect:intAP}
Since both ring current variations during storms and other (mainly auroral) current variations during strong substorms may produce significant GICs, we further performed similar SEAs for the $IntAp$ parameter, which (despite its own limitations, see section 2.2 and \cite{Kappenman2005}) is expected to roughly take into account the effects of both kinds of disturbances -- whereas $IntDst$ and $Int(d(\textit{SYM-H})/dt)$ only correspond to storm periods. However, due to the relatively low threshold $ap\geq 15$ (equivalent to $Kp\geq 3$) of integration used to calculate daily $IntAp$ levels, this new data series contained many more events (notably, many isolated substorms, sometimes outside of storms) than the previous $IntDst$ (storm) data set. As a result, requiring as before a 5-day period prior to events with $IntAp=0$ led to only a weak $IntAp$ maximum on Day 0, with a preceding $IntAp$ peak on Days $-10$ to $-5$ of comparable magnitude. Therefore, we changed our selection procedure, to consider only events with a peak $IntAp>1000$ nT$\cdot$hr and such that no similar event was present in the preceding 5 days. 

\begin{figure}
    \centering
    \resizebox{0.49\textwidth}{!}{\rotatebox{-90}{\includegraphics{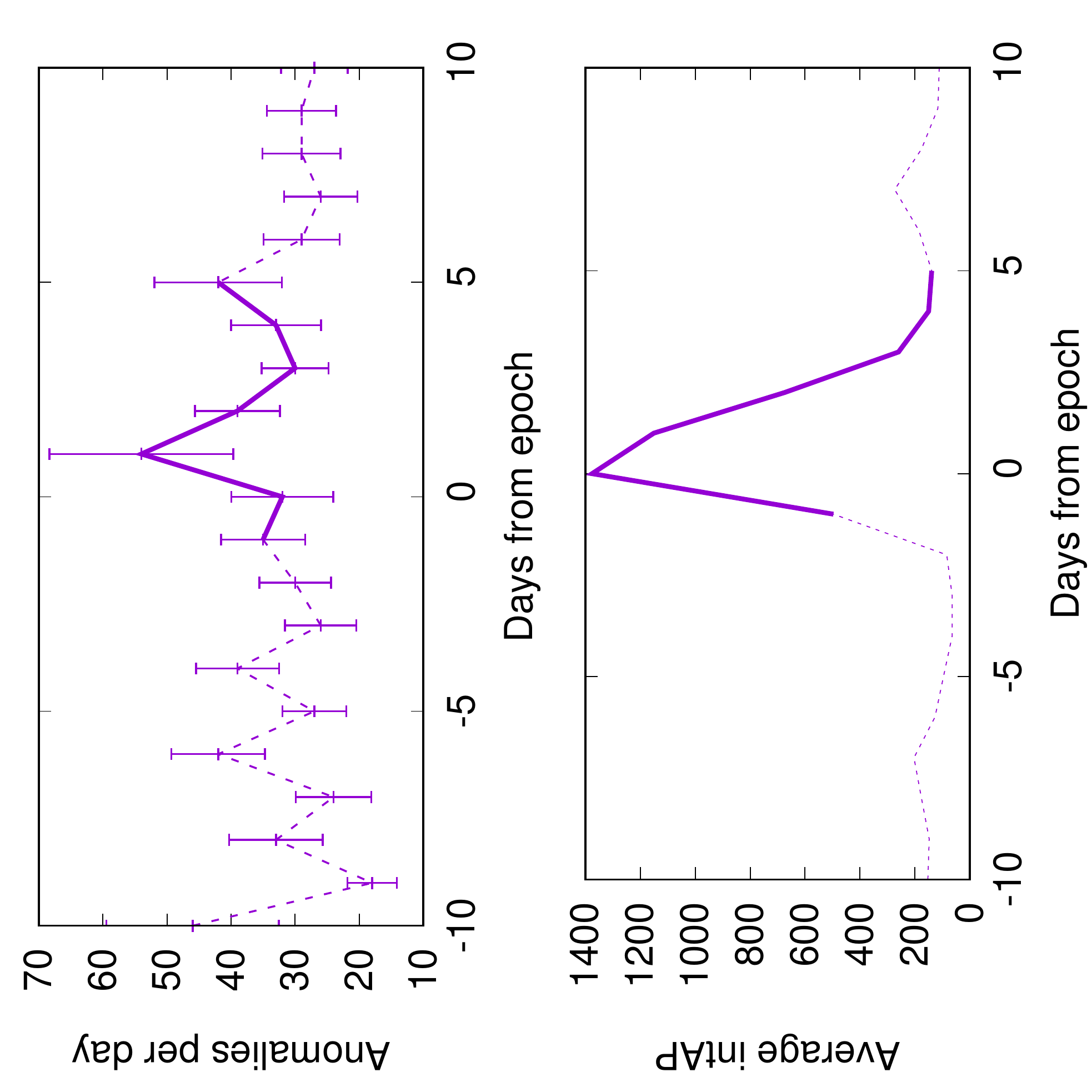}}}
    \resizebox{0.49\textwidth}{!}{\rotatebox{-90}{\includegraphics{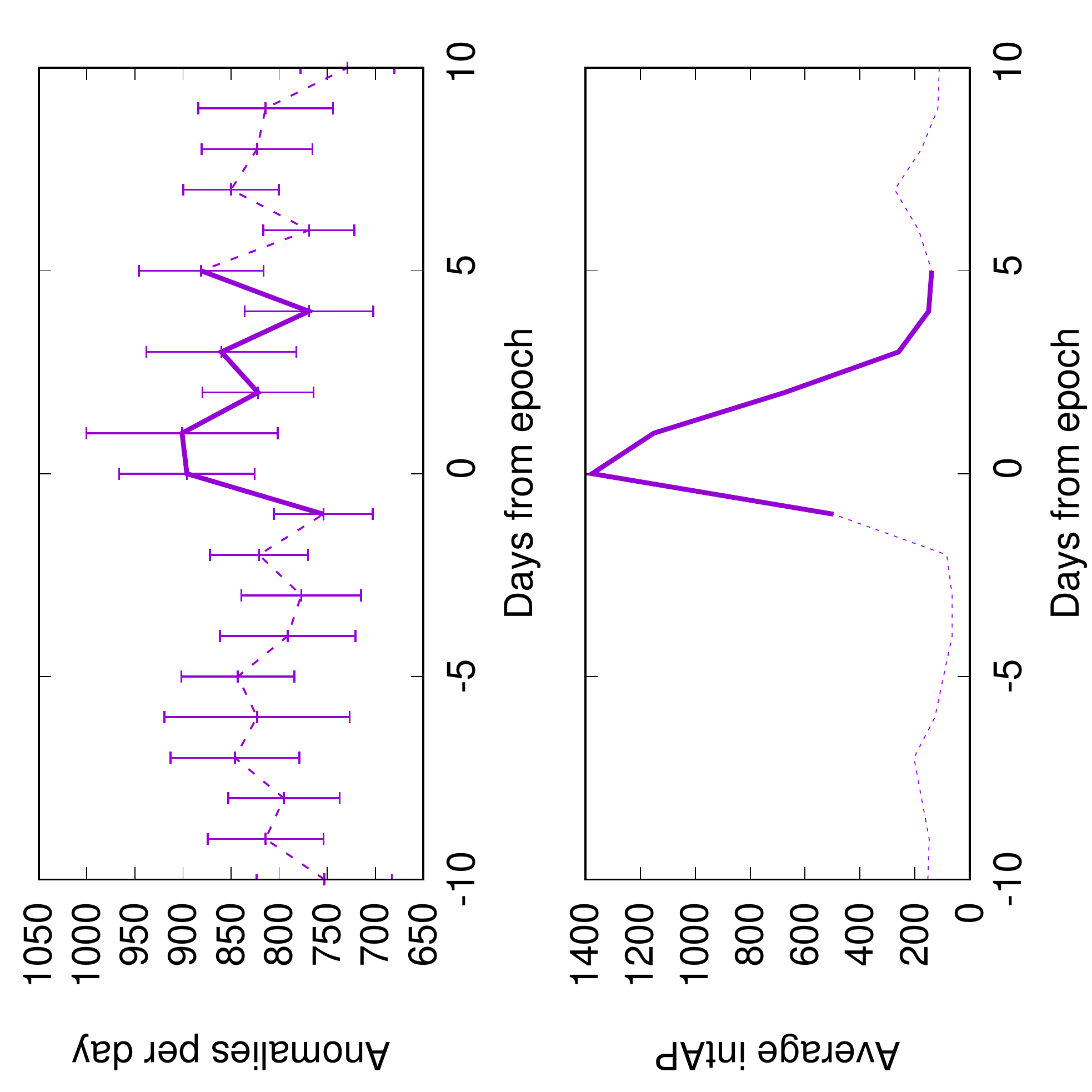}}}
    \caption{Plots of epoch-superposed subsets D11 (left) and D12 (right) of variations of the daily number of anomalies as a function of time, considering the $IntAp$ parameter. Solid lines indicate mean superposed daily rates of anomalies (upper row) or geomagnetic activity $IntAp$ (in nT$\cdot$hr) in the reference time series (bottom row) during Days $-1$ to $+5$ from the epoch (Day 0), whereas dashed lines show the same quantities for the remaining days. Error bars show the one-standard-deviation half-widths.}
    \label{fig:INTAP}
\end{figure}

The resulting SEAs displayed in Fig.~\ref{fig:INTAP} show that this new selection procedure produces a large peak $IntAp\sim 1400$ nT$\cdot$hr on Day 0 in the SEAs, with much lower levels on all 10 previous days, especially between Days $-6$ to $-2$. The daily number of anomalies is found to increase by a statistically very significant amount during the 5-day period following Day 0 as compared to the 5-day period preceding Day 0, for series D11 and D12 in Fig. \ref{fig:INTAP}, with corresponding $p$-values 0.03 and 0.007, respectively. There is a remarkable simultaneity between the peak of $IntAp$ and the peak of anomalies in the two SEAs with at most one day of delay. Moreover, such peaks of daily anomalies on Days 0 or $+1$ are consistently larger than all other daily values in the full 21-days SEAs. Such results therefore demonstrate the likely presence of nearly immediate effects of both storm-related and substorm-related geomagnetic disturbances on GICs and power lines (D11--D12) in the Czech power network. This is certainly due to the major impact of strong substorms on GICs, both during and outside geomagnetic storms.

There are also detectable increases of daily anomalies between 5-day periods before/after Day 0 for D8 (transformers, with a delay of $\sim 2$ days) and for D4 (power lines, immediate), but they are not statistically significant, with $p$-values $\simeq 0.25$ (see SEAs for all D1 to D12 series provided in Figs.~\ref{fig:D1intAP}--\ref{fig:D12intAP} in the online supplement). 

Besides, there is a statistically significant increase of anomalies for D10 (high and very high voltage electrical substations) with a $p$-value of 0.006, with a first peak of anomalies at Day $+1$ but a much delayed higher peak on Days $+4$ and $+5$. While power lines react immediately to GICs, high and very high voltage electrical substations, which comprise busbars, capacitors, or transformers, may indeed be affected but still continue to operate without registered problems until the cumulative damage reaches a sufficient level. A time lag of 3--5 days does not seem wholly unrealistic in this respect \citep{Kappenman2007, Wang2015}. 

It is worth noting that our previous analysis based on $IntDst$ did not show a statistically significant impact of storms for D10 (although the smallest $p$-value reached 0.08 in Table \ref{tab:pvalues}), contrary to the present analysis based on $IntAp$. This suggests that prolonged 2-3 day periods of repeated non-storm-time substorms or solar wind sudden impulses (SIs), taken into account by $IntAp$ but not by $IntDst$, could have a noticeable effect on some electrical substations.

\subsection{Auroral Current Effects: $IntAE$ parameter}

Next, we performed similar SEAs for the $IntAE$ parameter that provides a measure of cumulated high-latitude auroral current variations. An increased hourly auroral electrojet index $AE>150-250$ nT is one of the dominant manifestations of substorms, and many substorm studies rely on $AE$ to estimate the intensity of substorms, although $AE$ is not a specific measure of substorms \citep{Kamide1996, Tsurutani2004}. We compared the period of 5 {\it disturbed days} (with daily $IntAE>150$ nT$\cdot$hr) immediately following Day 0 (the day of peak $IntAE$) with the 5-day period immediately preceding Day 0 -- a preceding period of {\it nearly quiet days} (with daily $IntAE<30$ nT$\cdot$hr) especially selected to have such nearly zero $IntAE$ levels. This way, we can check the impact of {\it disturbed days} of strong $AE$ activity (often corresponding to substorms, occurring both during and outside storms) on power grid anomalies, as compared with {\it quiet days}. We also tried as before to consider shorter 3-day periods to help determine the best time lags between increases of anomalies and Day 0.

\begin{figure}
    \centering
     \resizebox{0.49\textwidth}{!}{\rotatebox{-90}{\includegraphics{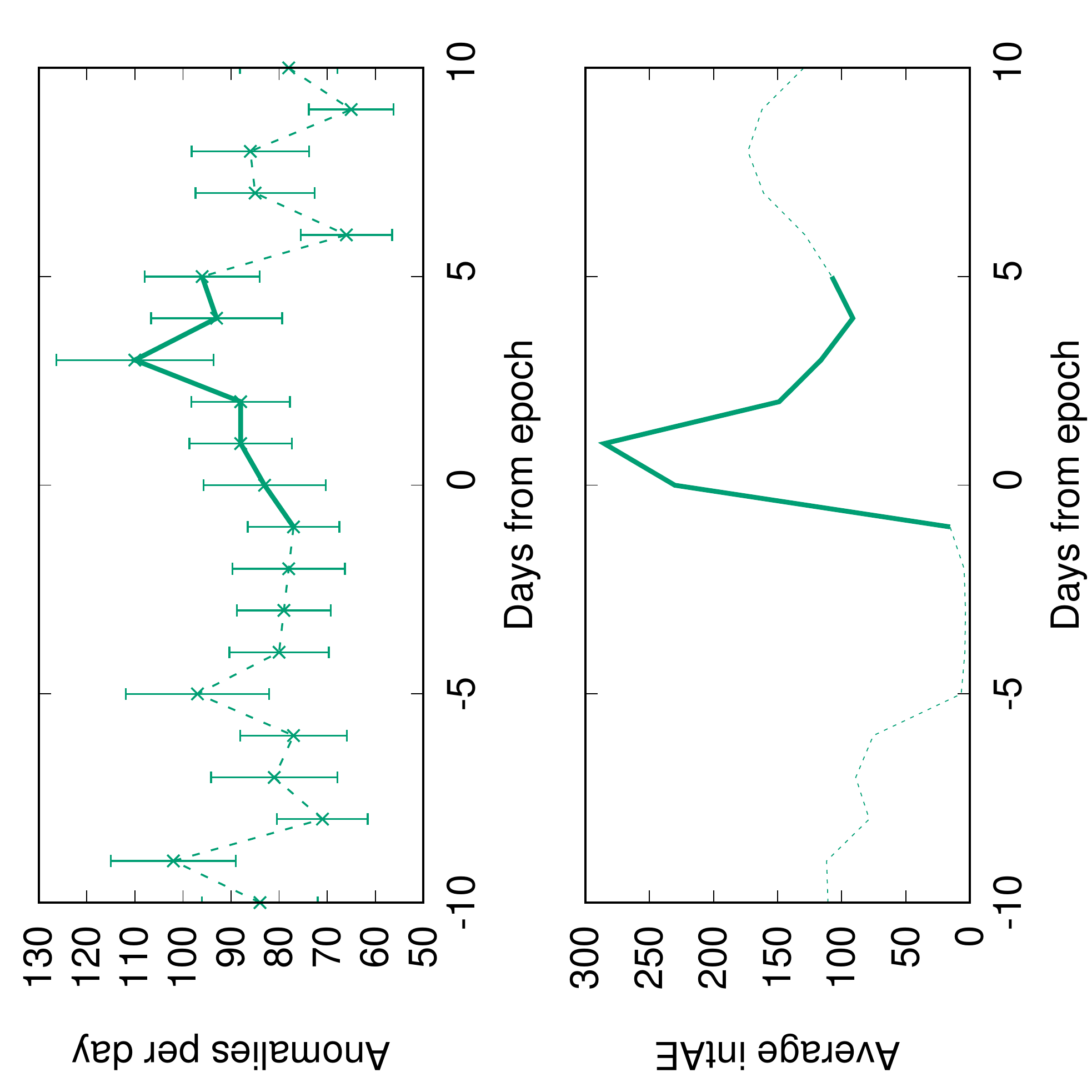}}}
    \caption{Epoch-superposed daily numbers of anomalies in the D11 series, considering the $IntAE$ parameter. Solid lines indicate superposed anomaly rates (upper panel) or $IntAE$ (in nT$\cdot$hr) in the reference time series (lower panel) during Days~$-1$ to $+5$ from the epoch (Day~0), whereas dashed lines show the same quantities for the remaining days. Error bars show one-standard-deviation half-widths. }
    \label{fig:IntAE}
\end{figure}

All the corresponding plots are given in Figs.~\ref{fig:D1intAE}--\ref{fig:D12intAE} in the Online attachment. In general, these results mostly agree with the $IntAp$ results. However, they are somewhat less statistically significant than the results obtained with all the preceding metrics, except for the D11-D12 (power lines) series. For D11, we find a statistically significant 15\% increase in the total number of anomalies after/before Day 0, with a $p$-value of 0.034 (see Fig. \ref{fig:IntAE}), while for D12 (power lines) the increase of anomalies is only 2.6\%, with a barely significant $p=0.055$. 

An important point is that these results based on $IntAE$ confirm the impact on power lines of auroral electrojet disturbances, often related to substorms. Nevertheless, these results also suggest that the $IntDst$, $IntAp$, or $Int(d(\textit{SYM-H})/dt)$ metrics may be slightly more appropriate than $IntAE$ for categorizing disturbed days leading to GIC effects at middle latitudes in the Czech power grid. This could stem from the higher latitudes of stations measuring $AE$ than for the mid-latitude $ap$ index: $IntAE$ may either take into account weak substorms that actually do not strongly affect middle latitudes, or it may under-estimate mid-latitude disturbances produced by large substorms \citep{Lockwood2019, Thomsen2004}. Alternatively, there could be some significant impacts of ring current variations on GICs at mid-latitudes, not taken into account in $IntAE$.

\section{Discussion}

In the SEAs, roughly $\approx 5-10$\% increases of the number of anomalies were often observed during the 5 most disturbed days as compared with the preceding 5 consecutive quiet days. However, it is important to note that such increases of anomalies were present during only the 5 most disturbed days among the 21-day total duration of each SEA. It is also unclear if there was any statistically significant increase of anomalies caused by the much weaker geomagnetic activity present during other days that did not fulfill the criteria for our SEA analysis. It is thus difficult to obtain a credible estimate of the total fraction of anomalies that could be directly related to geomagnetic effects. In our previous study \citep{Vybostokova2018}, the corresponding total number of anomalies attributable to variations of geomagnetic activity was also estimated as 1--4\%. Such values are consistent with results from a previous study of the impact of solar activity on the US electric power transmission network in 1992--2010, which showed that $\sim$ 4\% of the corresponding anomalies were likely attributable to strong geomagnetic activity and GICs \citep{schrijver2013disturbances}.

We also considered different parameter series, namely cumulative $IntDst$, $IntAp$, and $IntAE$ parameters integrated over the preceding 5 or 10 days, to evaluate the effects of a longer exposure to GICs on power-grid devices. The corresponding superposed epoch analysis did not yield statistically significant results. Without a proper event selection procedure and no integration limit, the SEAs were dominated by weak events, during which the effects were probably weak and did not emerge from the average rates of anomalies due to causes other than geomagnetic activity. SEAs were further performed separately for weak, moderate, and strong events, but this did not significantly improve the results. The most promising results in terms of magnitude of increase of anomalies during stronger activity were for D8, D10, and D12 for $IntDst$ (with lags of 1--3 days), and D8 and D11 for $IntAE$. 

Based on our analysis, it turns out that geomagnetic disturbances affected mostly the datasets registering anomalies on power lines. It is interesting to note that most of the power lines in D7, D11, and D12 are the power lines with distances between grounding points of the order of tens of kilometers. We also found significant delayed effects in the D8 dataset of high-voltage transformers. Although significant effects were observed in D4 during strong storms (see Fig.~S40), the distances between grounding points are of the order of hundreds of meters in this case, that is, much shorter than for the other power-line datasets. The topology of the network in D4 is also far more complex than in the other power-line datasets. It is unlikely that GICs induced in the D4 network could be responsible for the observed increase of anomaly rate after Day 0 in the corresponding SEA. Nevertheless, some detrimental currents could have entered the D4 network from nearby connected networks of other power companies and caused operational anomalies during strong events.

\section{Conclusions}

As noted by \cite{schrijver2013disturbances}, the selection of an appropriate geomagnetic parameter is very important when searching for correlations between anomalies recorded in human infrastructures and variations of geomagnetic activity. Here, we have presented results obtained by considering four different and complementary parameters of cumulative geomagnetic activity, namely the different storm-time $Int(d(\textit{SYM-H})/dt)$ and $IntDst$ low-latitude metrics tracking mainly ring current variations, the high-latitude $IntAE$ metric mainly tracking auroral current variations, and the mid-latitude $IntAp$ metric tracking both ring and auroral current variations -- all of which were integrated over geomagnetically disturbed periods. This allowed us to compare the cumulated number of anomalies observed in the Czech power grid during the corresponding disturbed days of high geomagnetic activity with the number of anomalies recorded during quiet days.

At the considered middle geomagnetic latitudes, our statistical analysis of $\sim10$ years of data has shown that space weather-related events affected mostly long power lines (D11, D12), probably due to a distortion of the electrical current waveform that eventually triggered a detrimental reaction of protective relays or disrupted other connected devices. However, significant and slightly more delayed (by $\sim1-2$ days) effects were also observed in high-voltage transformers. 

Both substorm-related disturbances and magnetic storms were found to have statistically significant impacts on the power grid network, since the four considered measures of disturbed days ($IntDst$, $Int(d(\textit{SYM-H})/dt)$, $IntAp$, and $IntAE$) led to more or less similar results -- although $IntAE$ was slightly less efficient. In addition, we found that considering moderate thresholds (neither too large nor too small) on time-integrated geomagnetic activity quantified by $IntDst$, $Int(d(\textit{SYM-H})/dt)$, or $IntAp$, produced the most statistically significant increases in anomaly rates, suggesting a non-negligible impact of moderate disturbances. These results are therefore consistent with a major impact of substorms, either inside or outside storms, on GICs at middle latitudes, together with a possible additional impact of ring current variations during storms.

It is worth noting that our study showed that in the 5-day period following the commencement of geomagnetic activity there is an approximately 5--10\% increase in the recorded power line and transformers anomalies in the Czech power grid, probably related to geomagnetic activity and GICs. Such values are consistent with previous results concerning the US power grid \citep{schrijver2013disturbances}. 
\cite{schrijver2014assessing} further found that for the US network, the 5\% stormiest days were apparently the most dangerous, with a 20\% increase of grid-related anomalies as compared to quiet periods. We similarly found that the days with a minimum $Dst<-50$ nT (roughly representing the $\approx 8$\% stormiest days, see \citealt{Gonzalez94}) had probably the strongest impact in the Czech power grid, leading to immediate or slightly delayed $\sim 5-20$\% increases of anomalies as compared to quiet periods.

\begin{acknowledgements}
M.\v{S} was supported by the institute research project RVO:67985815. We are grateful to power grid data providers for giving us an opportunity to exploit their logs of anomalies, namely to P.~Spurn\'y (\v{C}EPS), J.~Bro\v{z} and J.~Bu\v{r}i\v{c} (\v{C}EZ Distribuce), R.~Hanu\v{s} (PREdistribuce), and D.~Mezera and R.~B\'il\'y (E.ON Distribuce). The maintenance logs are considered strictly private by the power companies and are provided under non-disclosure agreements. We gratefully acknowledge the World Data Center in Kyoto and the Space Physics Data Facility (SPDF) at NASA Goddard Space Flight Center for the OMNI data at \url{http://omniweb.gsfc.nasa.gov} of $Dst$ and $\textit{SYM-H}$ geomagnetic indices used in this paper. {\bf Author contributions:} DM designed the study and provided processed geomagnetic data, K\v{Z} and TV wrote the processing code as parts of their student projects under the supervision of M\v{S}. M\v{S} performed the analysis. DM and M\v{S} interpreted the data and wrote the manuscript. All authors contributed to the final version of paper. 

\end{acknowledgements}





\Online

\begin{appendix} 
\section{Complete set of figures for all distributors}

\begin{figure}[!h]
    \resizebox{0.49\textwidth}{!}{\rotatebox{-90}{\includegraphics{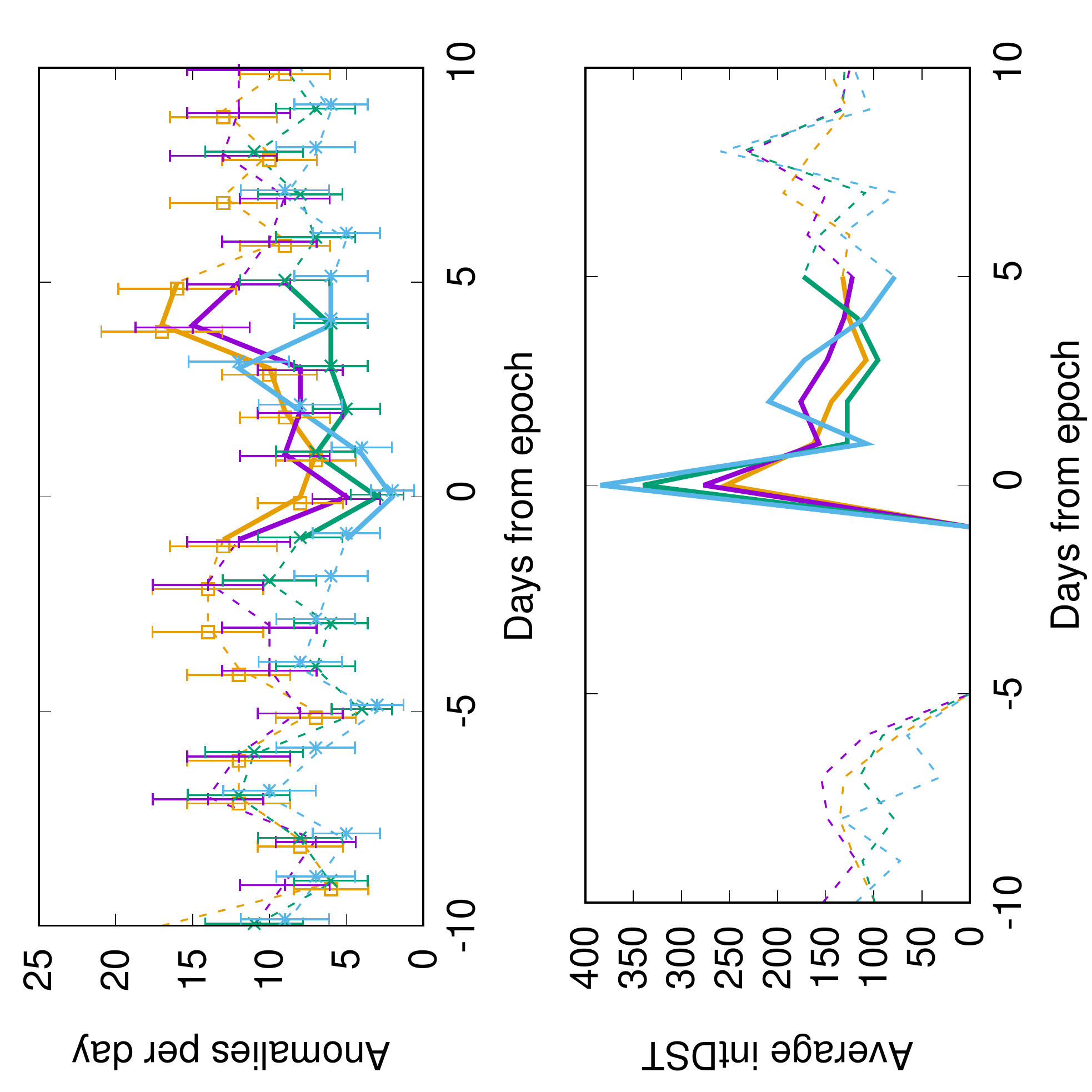}}}
    \resizebox{0.49\textwidth}{!}{\rotatebox{-90}{\includegraphics{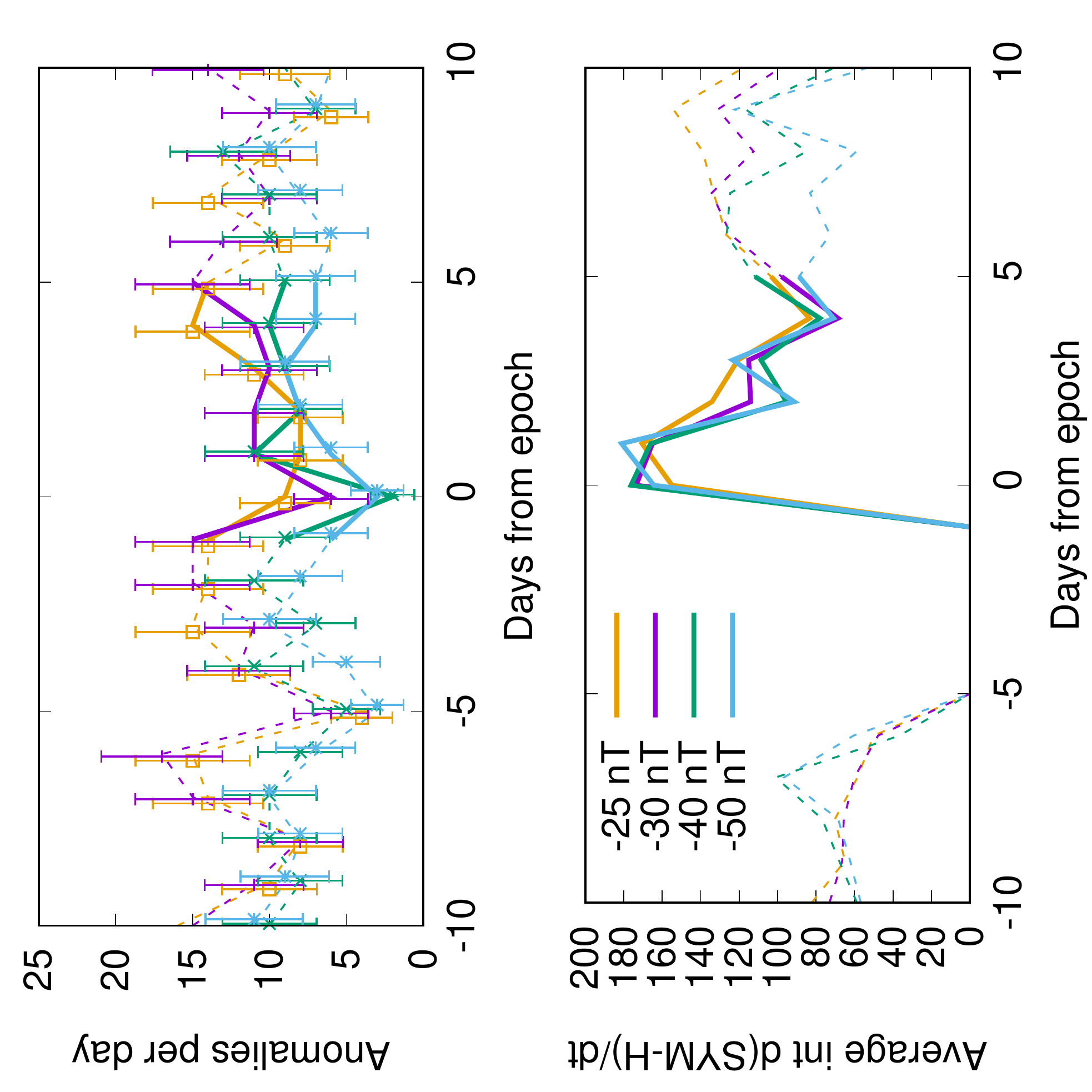}}}
    \caption{Plot of epoch-superposed D1 series considering $intDST$ with different thresholds (left) and $int(d(SYM-H)/dt)$ (right). The solid lines indicate the superposed anomaly rates (upper row) or reference time series (lower row) in the Days $-1$ to $+5$ from the epoch, whereas the dashed lines represent the same quantity for the remaining days.}
	\label{fig:D1seas}
\end{figure}

\foreach \n in {2,3,4,5,6,7,8,9,10,11}
{
\begin{figure}
    \resizebox{0.49\textwidth}{!}{\rotatebox{-90}{\includegraphics{app1a-\n.pdf}}}
    \resizebox{0.49\textwidth}{!}{\rotatebox{-90}{\includegraphics{app1b-\n.pdf}}}
    \caption{Same as Fig.~\ref{fig:D1seas} only for series D\n.}
\end{figure}
}

\begin{figure}
    \resizebox{0.49\textwidth}{!}{\rotatebox{-90}{\includegraphics{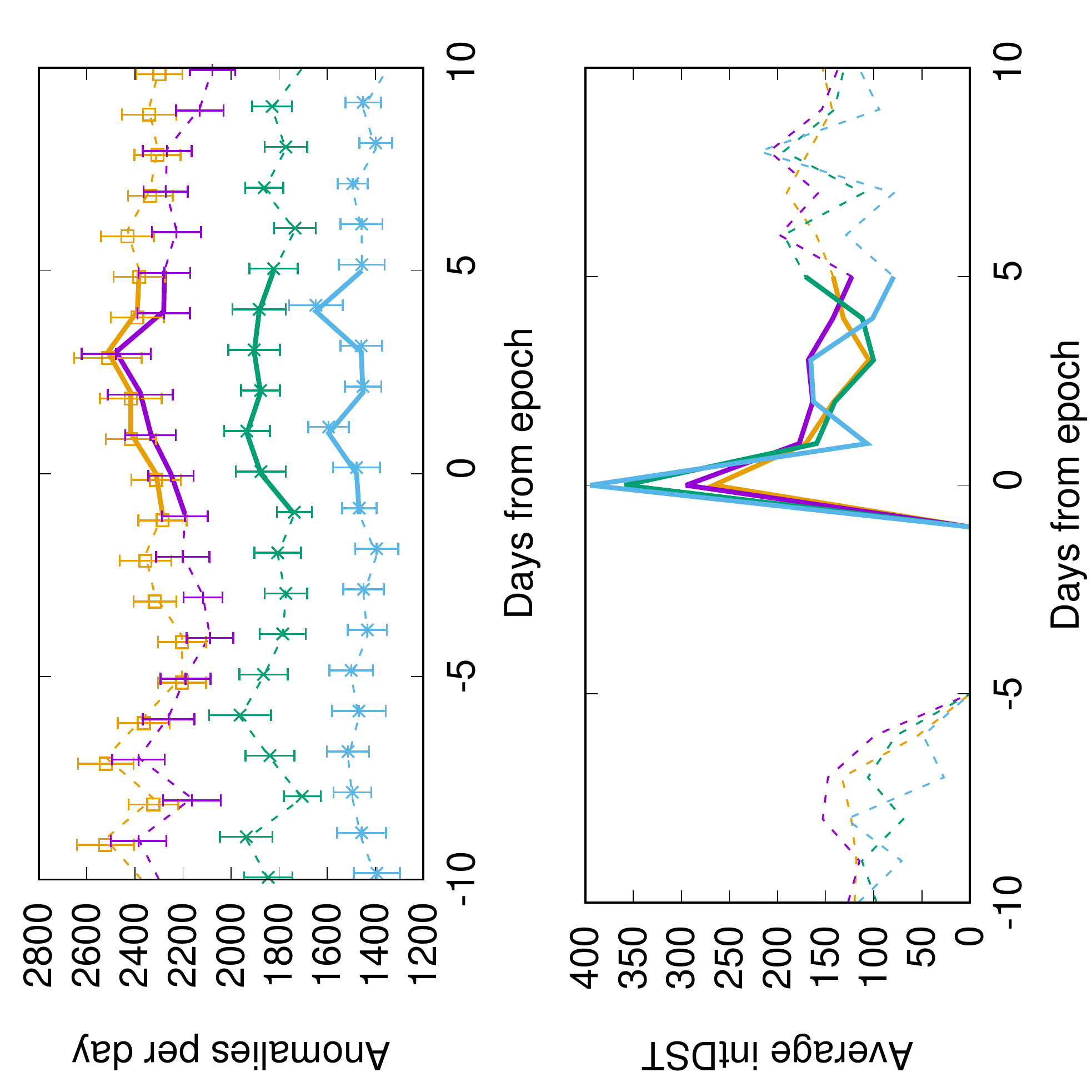}}}
    \resizebox{0.49\textwidth}{!}{\rotatebox{-90}{\includegraphics{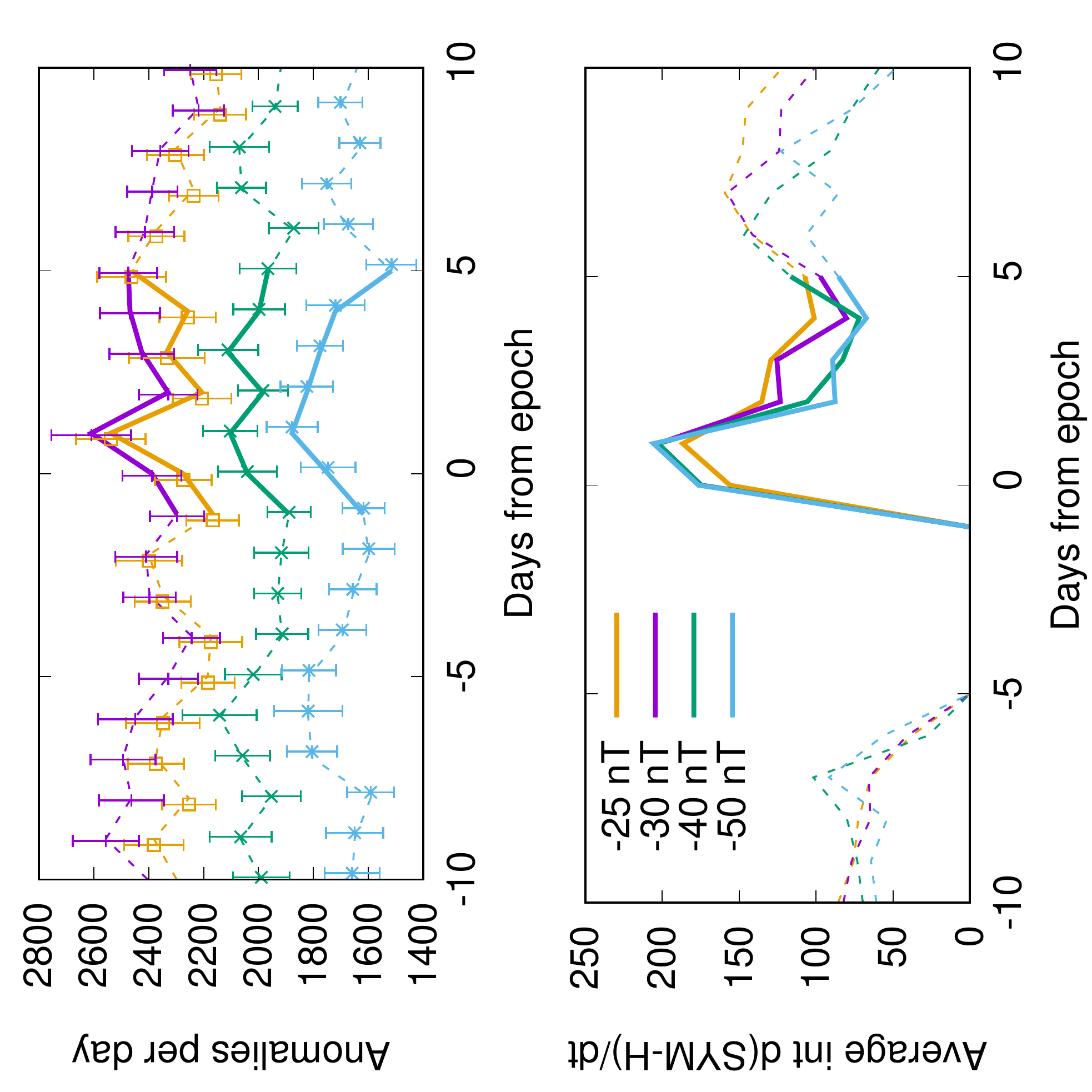}}}
    \caption{Same as Fig.~\ref{fig:D1seas} only for series D12.}
	\label{fig:D12seas}
\end{figure}
\clearpage

\begin{figure}
  \includegraphics[width=\textwidth]{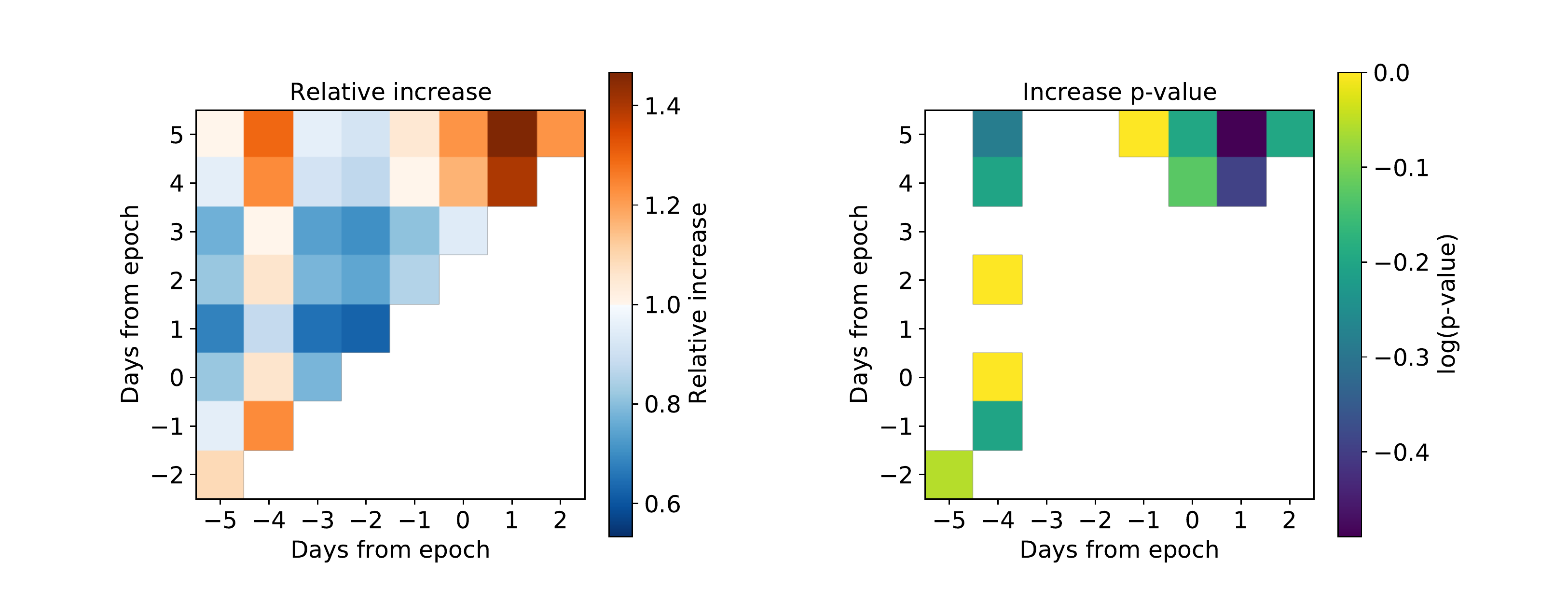}\\
  \includegraphics[width=\textwidth]{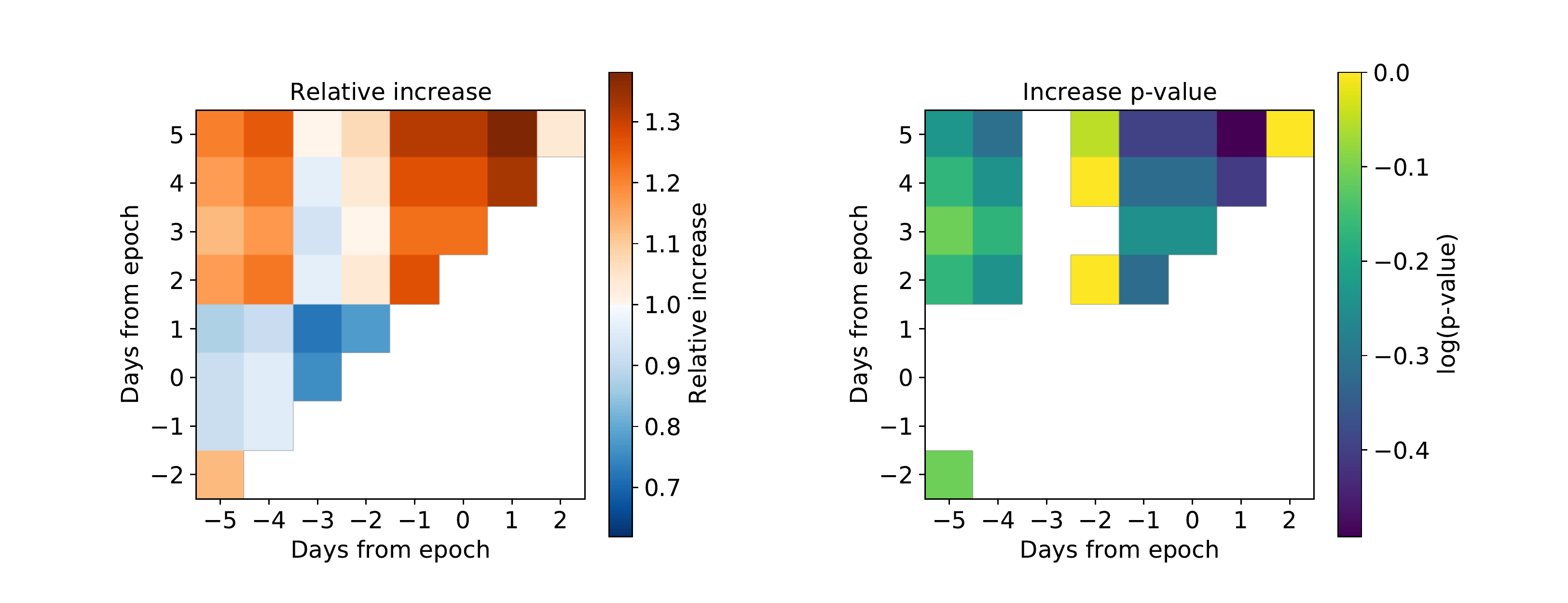}
  \caption{The map for D1 of increases (or decreases) of the number of anomalies as a function of the middle day of the first and second 3-day periods, together with maps of the corresponding $p$-values. The upper row is computed for $IntDST$ as the reference series, whereas the lower row is for $int(d(SYM-H)/dt)$ series. The $p$-values in the right column are evaluated only if there is an increase of the grid anomaly rates in the second 3-day period as compared to the first 3-day period. Note the logarithmic scale of the plotted $p$-values: $p=0.0055$ (the adopted level of statistical significance for individual bins) corresponds to $\log p=-2.26$. Statistically significant bins are indicated by white dots. Blank bins are indicated by the white colour.}
  \label{fig:D1ps}
\end{figure}

\foreach \n in {2,3,4,5,6,7,8,9,10,11}
{
\begin{figure}
  \includegraphics[width=\textwidth]{app2a-\n.pdf}\\
  \includegraphics[width=\textwidth]{app2b-\n.pdf}
  \caption{Same as Fig.~\ref{fig:D1ps} only for series D\n.}
\end{figure}
}

\begin{figure}
  \includegraphics[width=\textwidth]{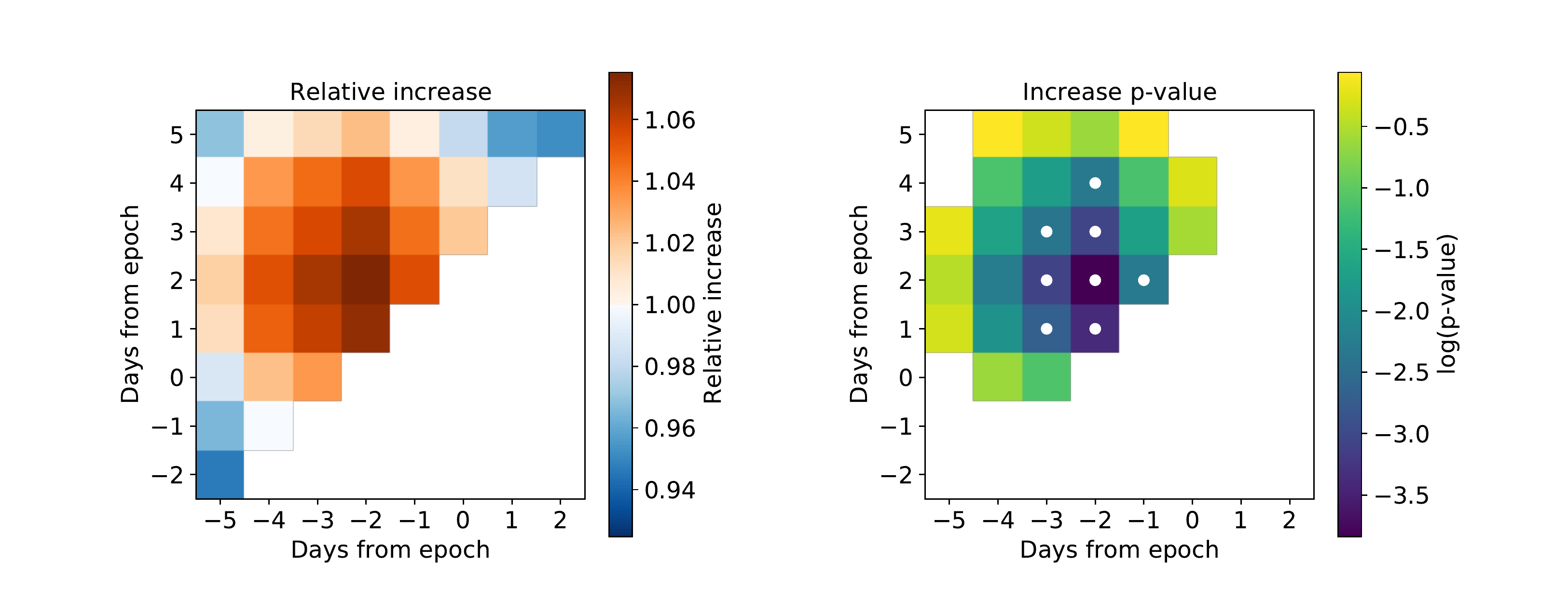}\\
  \includegraphics[width=\textwidth]{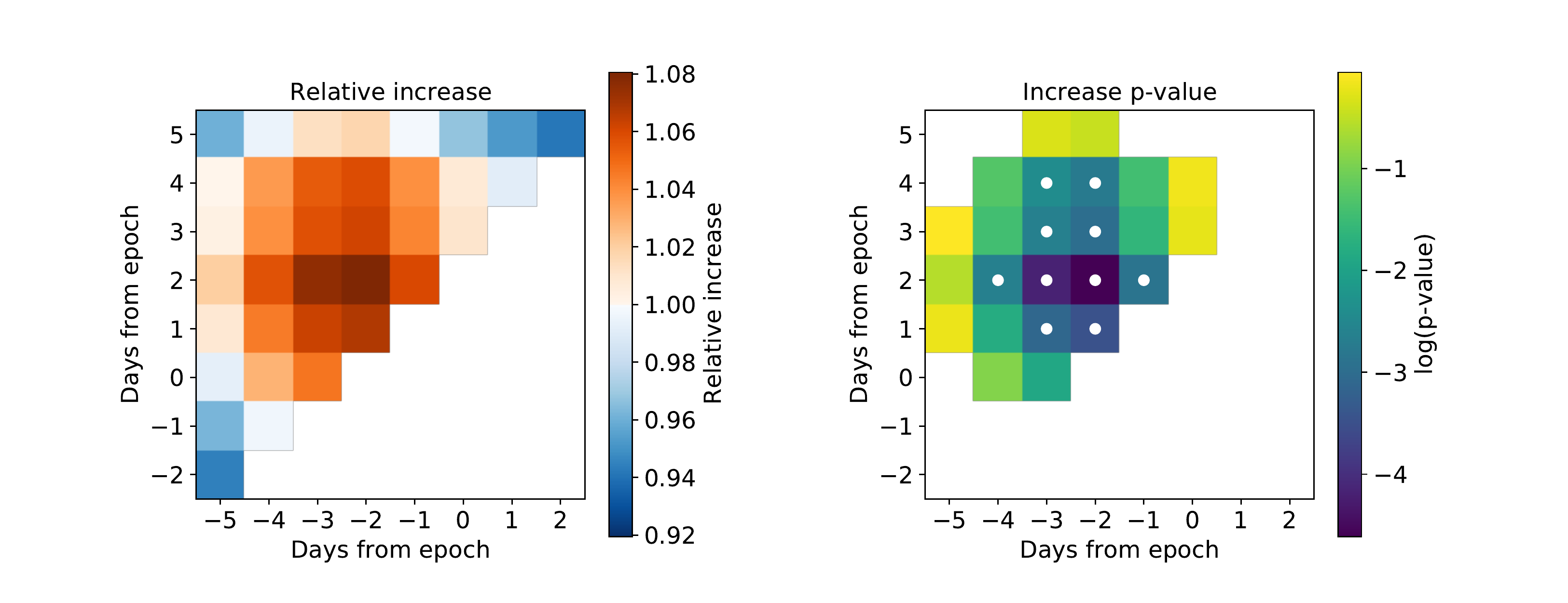}
  \caption{Same as Fig.~\ref{fig:D1ps} only for series D12.}
  \label{fig:D12ps}
\end{figure}

\clearpage

\begin{figure}
  \centering
  \resizebox{0.49\textwidth}{!}{\rotatebox{-90}{\includegraphics{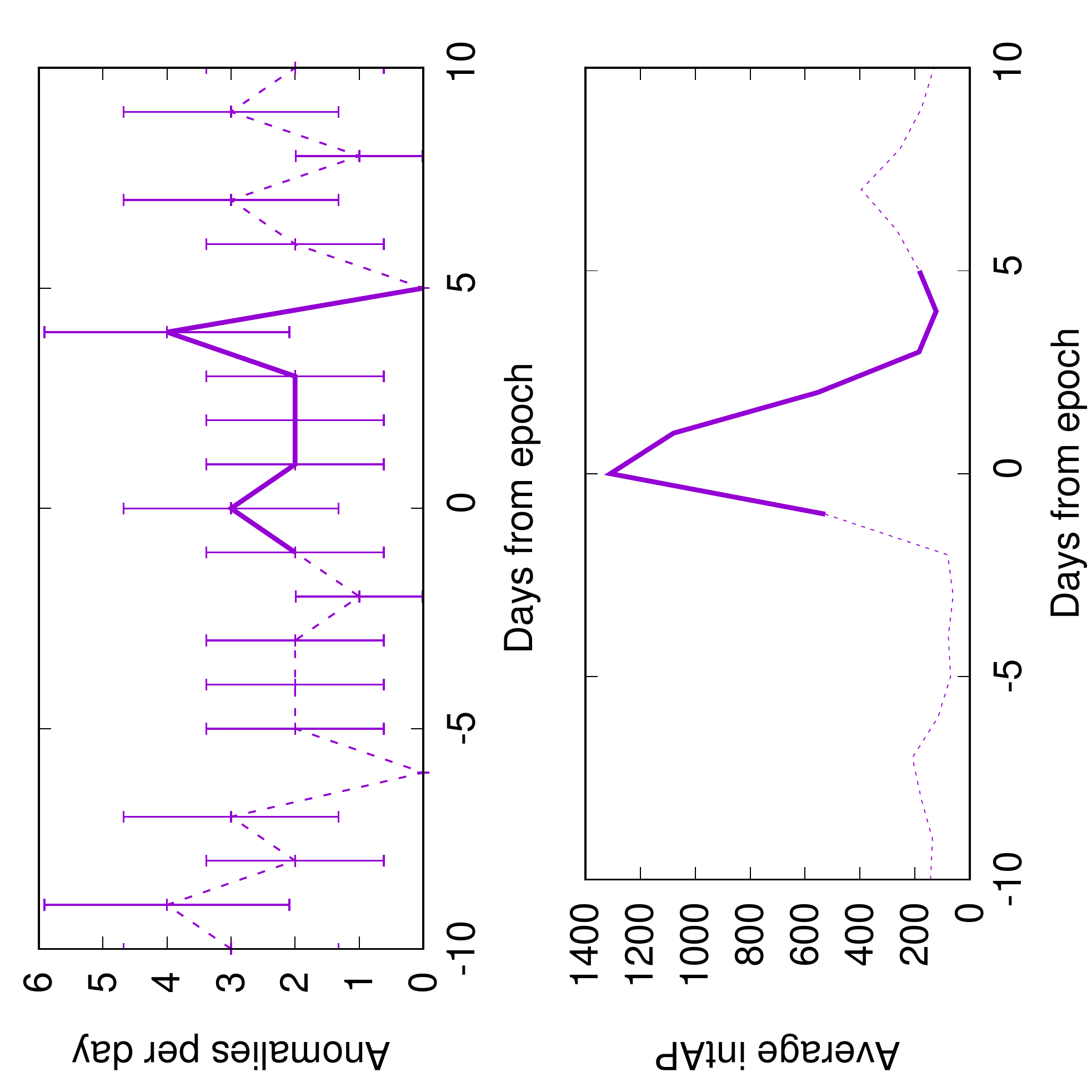}}}
  \caption{Plot of epoch-superposed D1 of increases (or decreases) of the daily number of anomalies as a function of time, 
  considering the $IntAp$ parameter. Solid lines indicate superposed daily rates of anomalies (upper row) or geomagnetic 
  activity $IntAp$ in the reference time series (bottom row) during Days $-1$ to $+5$ from the epoch (Day 0), 
  whereas dashed lines show the same quantities for the remaining days.}
  \label{fig:D1intAP}
\end{figure}

\foreach \n in {2,3,4,5,6,7,8,9,10,11}
{
\begin{figure}
  \centering
  \resizebox{0.49\textwidth}{!}{\rotatebox{-90}{\includegraphics{app3-\n.pdf}}}
  \caption{Same as Fig.~\ref{fig:D1intAP} only for series D\n.}
\end{figure}
}

\begin{figure}
  \centering
  \resizebox{0.49\textwidth}{!}{\rotatebox{-90}{\includegraphics{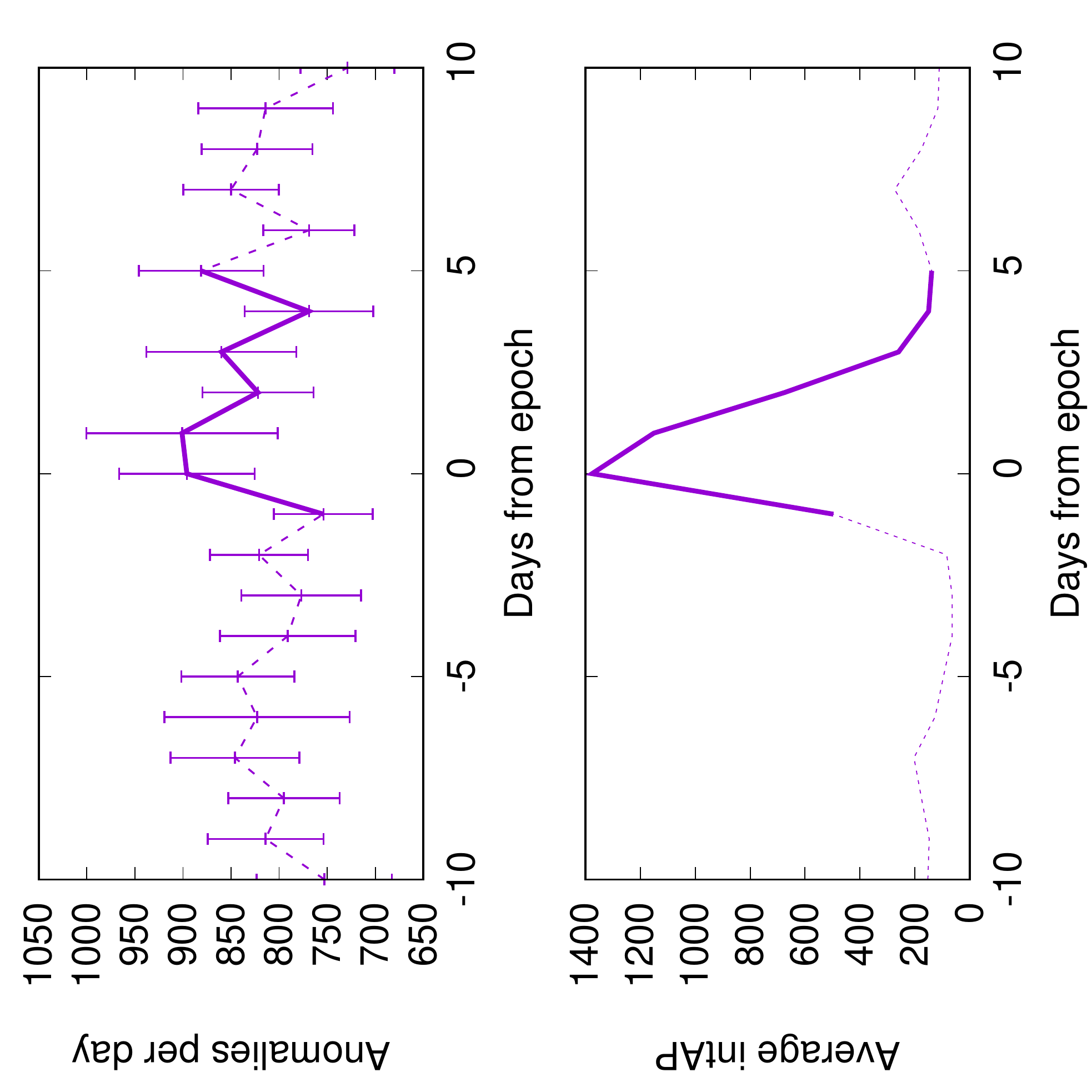}}}
  \caption{Same as Fig.~\ref{fig:D1intAP} only for series D12.}
  \label{fig:D12intAP}
\end{figure}

\clearpage

\begin{figure}
  \centering
  \resizebox{0.49\textwidth}{!}{\rotatebox{-90}{\includegraphics{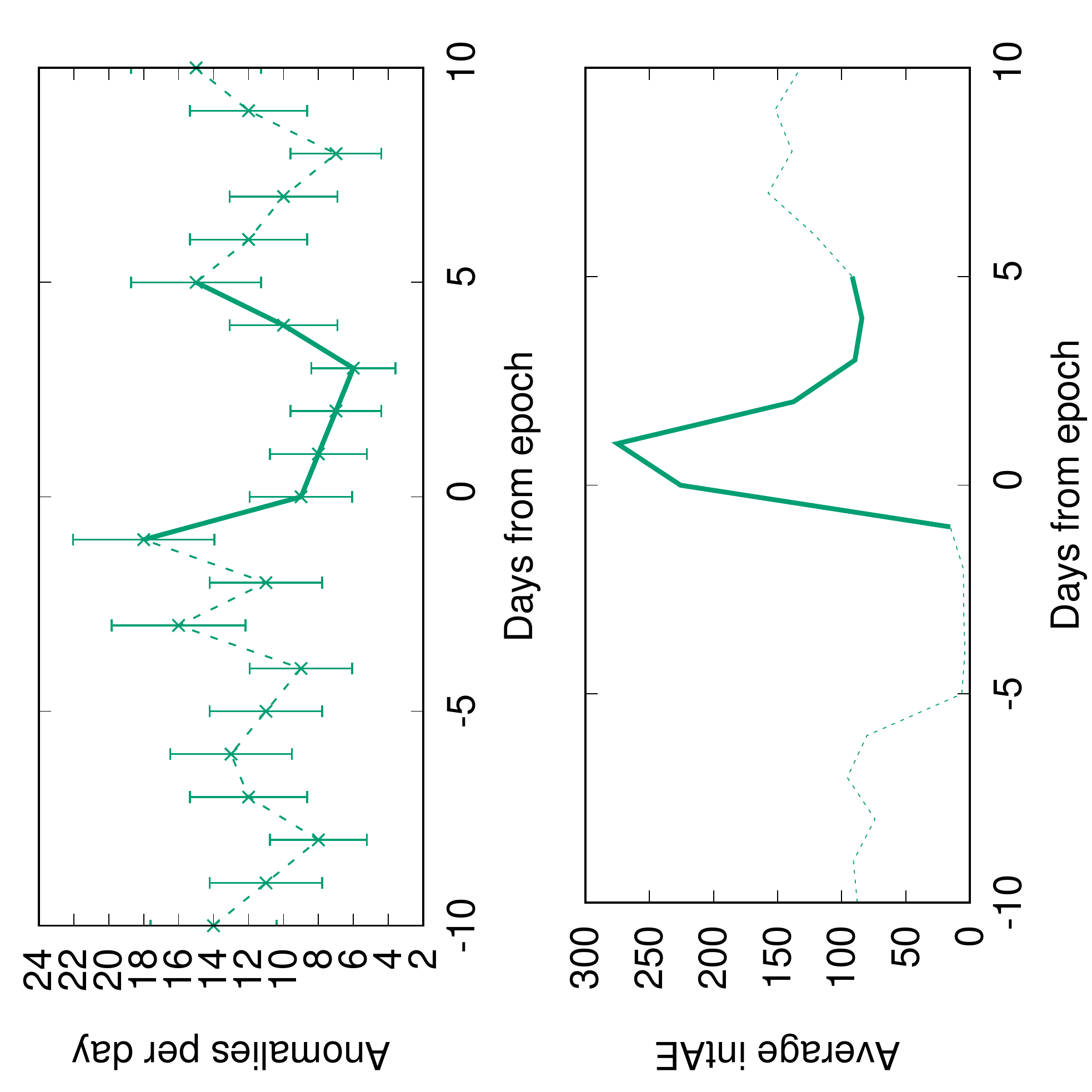}}}
  \caption{Plot of epoch-superposed D1 of increases (or decreases) of the daily number of anomalies as a function of time, 
  considering the $IntAE$ parameter. Solid lines indicate superposed daily rates of anomalies (upper row) or geomagnetic 
  activity $IntAE$ in the reference time series (bottom row) during Days $-1$ to $+5$ from the epoch (Day 0), 
  whereas dashed lines show the same quantities for the remaining days.}
  \label{fig:D1intAE}
\end{figure}

\foreach \n in {2,3,4,5,6,7,8,9,10,11}
{
\begin{figure}
  \centering
  \resizebox{0.49\textwidth}{!}{\rotatebox{-90}{\includegraphics{app4-\n.pdf}}}
  \caption{Same as Fig.~\ref{fig:D1intAE} only for series D\n.}
\end{figure}
}

\begin{figure}
  \centering
  \resizebox{0.49\textwidth}{!}{\rotatebox{-90}{\includegraphics{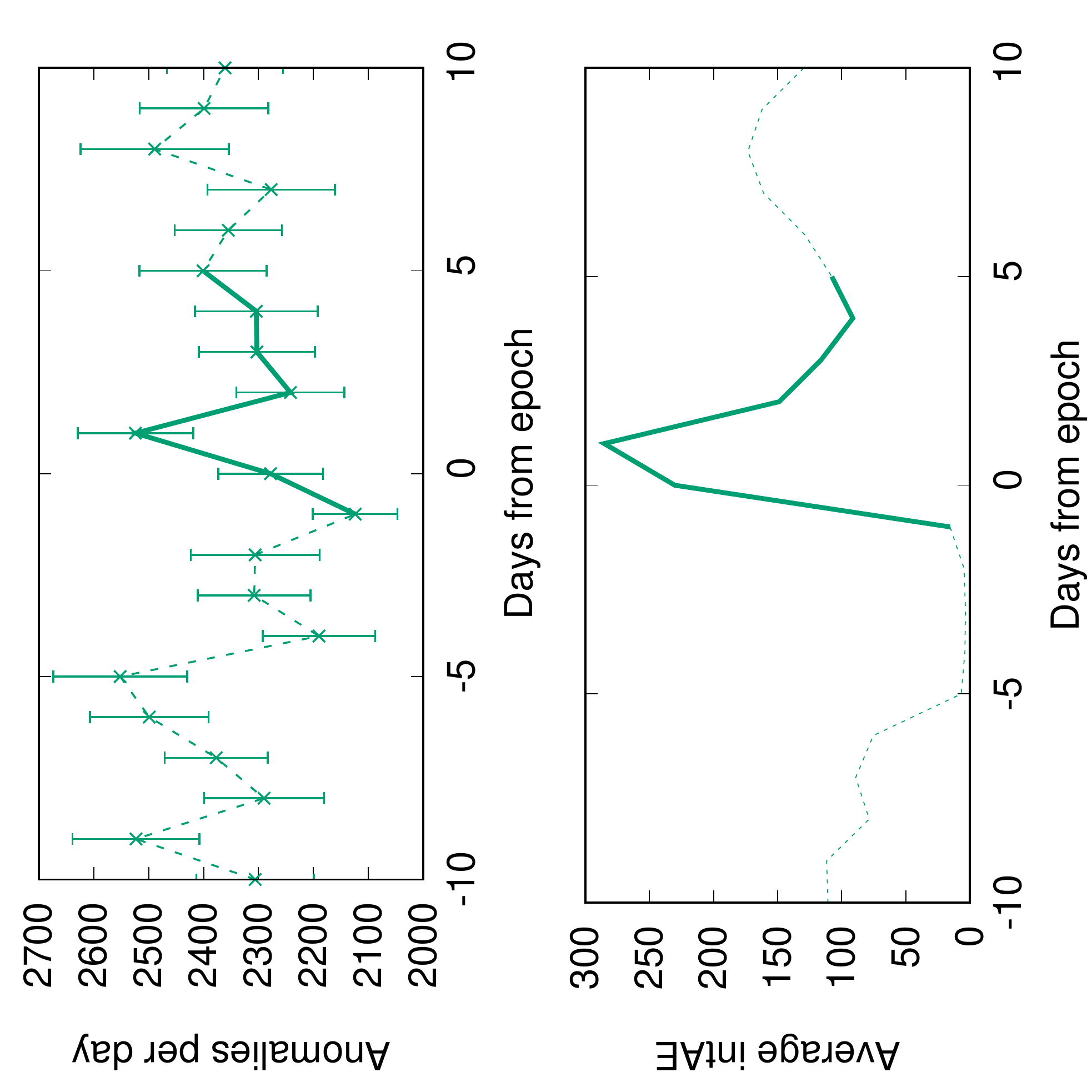}}}
  \caption{Same as Fig.~\ref{fig:D1intAE} only for series D12.}
  \label{fig:D12intAE}
\end{figure}

\end{appendix}

\end{document}